\documentclass[english,aps,prc,nofootinbib,superscriptaddress,twocolumn,letterpaper,floatfix]{revtex4}
\usepackage[latin9]{inputenc}
\usepackage{hyperref}
\hypersetup{
  colorlinks=true,
  linkcolor=blue,
  citecolor=cyan,
}
\usepackage{breakurl}
\usepackage{graphicx}
\usepackage{amssymb,amsmath}
\usepackage[usenames]{color}
\usepackage{times}
\usepackage{comment}
\usepackage{booktabs}

\newcommand\snn{\sqrt{s_\text{NN}}}

\newcommand{\trento}{T\raisebox{-0.5ex}{R}ENTo-3D}

\allowdisplaybreaks


\begin{document}

\title{Hyperon polarization in isobaric Zr+Zr collisions at $\snn=200$~GeV: \trento{} +
       CLVisc with an initial longitudinal flow gradient}

\author{Ze-Fang Jiang}
\email{jiangzf@mails.ccnu.edu.cn}
\affiliation{Department of Physics and Electronic-Information Engineering,
             Hubei Engineering University, Xiaogan, Hubei, 432000, China}
\affiliation{Institute of Particle Physics and Key Laboratory of Quark and
             Lepton Physics (MOE), Central China Normal University, Wuhan,
             Hubei, 430079, China}

\author{Xiang Fan}
\affiliation{Institute of Particle Physics and Key Laboratory of Quark and
             Lepton Physics (MOE), Central China Normal University, Wuhan,
             Hubei, 430079, China}
\affiliation{Faculty of Physics, Bielefeld University, D-33615 Bielefeld, Germany}

\author{Jing Jing}
\affiliation{Department of Physics and Electronic-Information Engineering,
             Hubei Engineering University, Xiaogan, Hubei, 432000, China}

\begin{abstract}
We present a theoretical study of global and azimuthal-angle-dependent $\Lambda$
hyperon polarization in isobaric $^{96}_{40}$Zr+$^{96}_{40}$Zr collisions at
$\sqrt{s_{NN}}=200$~GeV using the \trento\ initial condition model coupled to the
(3+1)-D viscous hydrodynamic model CLVisc. A longitudinal flow velocity gradient,
controlled by $f_v$, is introduced into \trento\ for the first time, providing an
essential source of initial vorticity in this symmetric isobaric system. Within the
isothermal polarization framework, the model provides a simultaneous description of
STAR measurements of the global polarization $-P^{y}$ (centrality, $p_T$, and $\eta$
dependences) and the azimuthal modulation coefficients $P_{y,\mathrm{c0}}$ and
$P_{y,\mathrm{c2}}$. The $p_T$ dependence reflects the competition between thermal
vorticity and shear contributions: the thermal term decreases with $p_T$, while the
shear term rises and increasingly shapes the curvature of the total polarization. In
this decomposition, $P_{y,\mathrm{c2}}$ is dominantly shear-driven and serves as a
clean probe of shear-induced polarization. Scans of $f_v$, $k_T$, and nuclear
structure provide complementary constraints on the initial state, while the
bulk-viscosity dependence is also examined; the five nuclear structure
configurations from the STAR isobar blind analysis yield nearly indistinguishable
polarization. For $P_z$, the isothermal scenario captures the
azimuthal modulation but overpredicts the high-$p_T$ modulation amplitude, and
comparison with the standard thermal treatment shows that neither scenario achieves a
unified description of all observables.
\end{abstract}

\date{\today}
\maketitle

\section{Introduction}
\label{emsection1}

The strongly coupled quark-gluon plasma (QGP) created in relativistic heavy-ion
collisions at the Relativistic Heavy-Ion Collider (RHIC) and the Large Hadron Collider
(LHC) has been established as a nearly perfect fluid with remarkably low shear
viscosity~\cite{Heinz:2013th,Busza:2018rrf,Bernhard:2019bmu}. In non-central
collisions, the substantial orbital angular momentum generates the most vortical fluid
ever observed in nature, whose existence is confirmed by the global polarization of
emitted hyperons via spin-vorticity coupling~\cite{Liang:2004ph,Voloshin:2004ha,
Becattini:2007sr,
Becattini:2013fla,STAR:2017ckg,Becattini:2017gcx,STAR:2018gyt,ALICE:2021pzu}. This
discovery has opened an entirely new direction in heavy-ion physics: spin polarization
provides direct access to the vortical and shear structure of the expanding medium,
complementary to the well-established collective flow
observables~\cite{Karpenko:2016jyx,Xia:2018tes,Sun:2017xhx,Li:2017slc,Voloshin:2017kqp,Ivanov:2020wak,Alzhrani:2022dpi,STAR:2025dgs}.

Significant theoretical progress has been made in understanding the mechanisms
underlying hyperon polarization. Hydrodynamic and transport models have identified the
thermal vorticity tensor $\varpi_{\alpha\beta}\propto
\partial_{[\alpha}\left(u_{\beta]}/T\right)$ as the primary source of the global out-of-plane
polarization $-P^{y}$, while the symmetric shear tensor $\xi_{\alpha\beta}$ contributes
importantly to the local polarization and its azimuthal
modulation~\cite{Fu:2021pok,Becattini:2021suc,Becattini:2021iol,Liu:2020dxg,Fu:2020oxj,Yi:2021ryh,Wu:2022mkr,Dey:2026epy}.
However, several challenges remain. First, the proper treatment of
temperature-gradient terms in the spin polarization formula remains under debate,
since the isothermal and standard thermal treatments give different predictions,
particularly for $P_z$ (the so-called ``sign
puzzle'')~\cite{Florkowski:2019voj,Becattini:2021suc,Palermo:2024tza}. Second, the
three-dimensional structure of the initial fireball geometry and the longitudinal flow
velocity profile, which jointly determine the fluid vorticity, are not well constrained
by existing data~\cite{Shen:2020jwv,Ryu:2021lnx,Jiang:2023vxp}. Third, while
polarization has been studied extensively in symmetric Au+Au and Pb+Pb
systems~\cite{STAR:2018gyt,ALICE:2021pzu}, the isobaric collisions of
$^{96}$Ru+$^{96}$Ru and $^{96}$Zr+$^{96}$Zr at $\snn=200$~GeV, performed by the STAR
Collaboration to constrain the chiral magnetic effect
(CME~\cite{Kharzeev:2020jxw,Fukushima:2008xe}) and nuclear
structure~\cite{STAR:2021mii}, provide a controlled setting for testing whether
hyperon polarization is sensitive to isobaric nuclear-structure variations and to the
three-dimensional initial condition.

The STAR Collaboration has recently released measurements of both the global and
azimuthal-angle-dependent polarization of $\Lambda$ hyperons in isobaric
collisions~\cite{STAR:2025dgs,STAR:2023eck}. These data, which include the centrality
and transverse momentum dependences of $-P^{y}$, the Fourier coefficients
$P_{y,\mathrm{c0}}$ and $P_{y,\mathrm{c2}}$ characterizing the azimuthal modulation,
and the longitudinal polarization $P_z$, provide a rich set of constraints for
theoretical models. To date, however, a comprehensive theoretical description of these
observables within a single, self-consistent framework has been lacking.

In this work, we perform a systematic study of $\Lambda$ hyperon polarization in Zr+Zr
collisions at $\snn=200$~GeV using the \trento\ initial condition
model~\cite{Soeder:2023vdn} coupled to the (3+1)-D viscous hydrodynamic code
CLVisc~\cite{Pang:2018zzo,Wu:2021fjf}. We extend \trento\ by incorporating an initial
longitudinal flow velocity gradient via a tunable parameter $f_v$~\cite{Ryu:2021lnx,
Jiang:2023vxp}, thereby moving beyond the Bjorken approximation. This provides a source
of longitudinal vorticity essential for polarization in symmetric isobaric systems.
Within the isothermal polarization
scenario~\cite{Yi:2024kwu,Becattini:2021suc,Becattini:2021iol,Palermo:2024tza}, we
systematically investigate the sensitivity of polarization to the initial longitudinal
flow gradient $f_v$, the fragmentation-induced geometric tilt controlled by $k_T$, the
five nuclear structure configurations~\cite{Jia:2022qgl,STAR:2021mii}, and the bulk
viscosity $\zeta/s$, and compare the isothermal and standard thermal treatments.

With the longitudinal flow gradient $f_v=0.10$ and $k_T=0.33$~GeV, the framework
provides a simultaneous description of the STAR measurements of $-P^{y}$ (its
centrality, $p_T$, and $\eta$ dependences), together with the azimuthal coefficients
$P_{y,\mathrm{c0}}$ and $P_{y,\mathrm{c2}}$. The $p_T$ dependence of $-P^{y}$ reflects
the competition between thermal vorticity and shear contributions, with the Fourier
coefficients cleanly separating the two: $P_{y,\mathrm{c0}}$ is vorticity-dominated,
while $P_{y,\mathrm{c2}}$ is shear-driven. Systematic scans reveal that $f_v$ primarily
controls the overall magnitude, $k_T$ affects both the low-$p_T$ level and the $\eta$
profile, while nuclear structure has negligible impact. For the longitudinal
polarization $P_z$, the isothermal scenario captures the azimuthal modulation but
overpredicts the high-$p_T$ $\langle P_z\sin[2(\phi-\Psi_2)]\rangle$, and neither the
isothermal nor the standard thermal treatment provides a unified description of all
observables, highlighting current limitations and motivating further theoretical
development.

The remainder of this paper is organized as follows. Section~\ref{v1section2} presents
the theoretical framework, including the \trento\ initialization with the longitudinal
flow extension, the CLVisc hydrodynamic model with bulk viscosity, and the isothermal
polarization formalism. Section~\ref{section3} contains our numerical results: baseline
observables and global polarization (\ref{sec:3-1}), sensitivity to $f_v$, $k_T$, and
nuclear structure (\ref{sec:3-2}), azimuthal-angle-dependent polarization coefficients
(\ref{sec:3-3}), longitudinal polarization (\ref{sec:3-4}), and the comparison between
isothermal and thermal scenarios (\ref{sec:3-5}). A summary and outlook are given in
Sec.~\ref{section4}.


\section{Theoretical framework}
\label{v1section2}

\subsection{\texorpdfstring{\trento\ initialization with longitudinal flow
gradient}{TRENTo-3D initialization with longitudinal flow gradient}}
\label{sec:trento3d_detail}

We study isobaric $^{96}_{40}$Zr+$^{96}_{40}$Zr collisions at $\snn=200$~GeV. The
three-dimensional initial condition for the hydrodynamic evolution of the QGP medium is
generated using the \trento\ model~\cite{Soeder:2023vdn}, which has been systematically
calibrated to describe charged particle rapidity distributions across RHIC and LHC
energies.

A key input to the \trento\ framework is the nuclear geometry of the colliding species.
For $^{96}$Zr, we adopt a deformed Woods-Saxon distribution~\cite{Loizides:2017ack}:
\begin{equation}
\rho(r,\theta,\phi) = \frac{\rho_{0}}{1 + \exp\bigl[(r - R(\theta,\phi))/a_{0}\bigr]},
\label{eq:WS_deformed}
\end{equation}
where the angle-dependent radius is expanded in spherical harmonics as
\begin{equation}
R(\theta,\phi) = R_{0}\bigl[1 + \beta_{2}Y_{2}^{0}(\theta) +
\beta_{3}Y_{3}^{0}(\theta)\bigr],
\label{eq:R_deformed}
\end{equation}
with $R_{0}=5.02$~fm, $a_{0}=0.52$~fm, $\beta_{2}=0.06$, and $\beta_{3}=0.20$. These
parameters are taken from Ref.~\cite{Jia:2022qgl} (Case~5 therein) and have been
validated through comparisons of anisotropic flow and multiplicity distributions in
Ru+Ru and Zr+Zr collisions. The non-zero octupole deformation $\beta_{3}=0.20$ is a
distinctive feature of the $^{96}$Zr nucleus that can influence the initial geometry of
the QGP and, consequently, the vorticity field responsible for hyperon polarization. To
assess the sensitivity of our polarization results to nuclear structure uncertainties,
we also perform calculations using the alternative parameter sets originally proposed
for $^{96}$Ru in the analysis of Ref.~\cite{Jia:2022qgl}; these comparisons are
presented in Sec.~\ref{subsubsec-3-3}.

\begin{figure*}[tbp!]
\begin{center}
\includegraphics[width=0.85\linewidth]{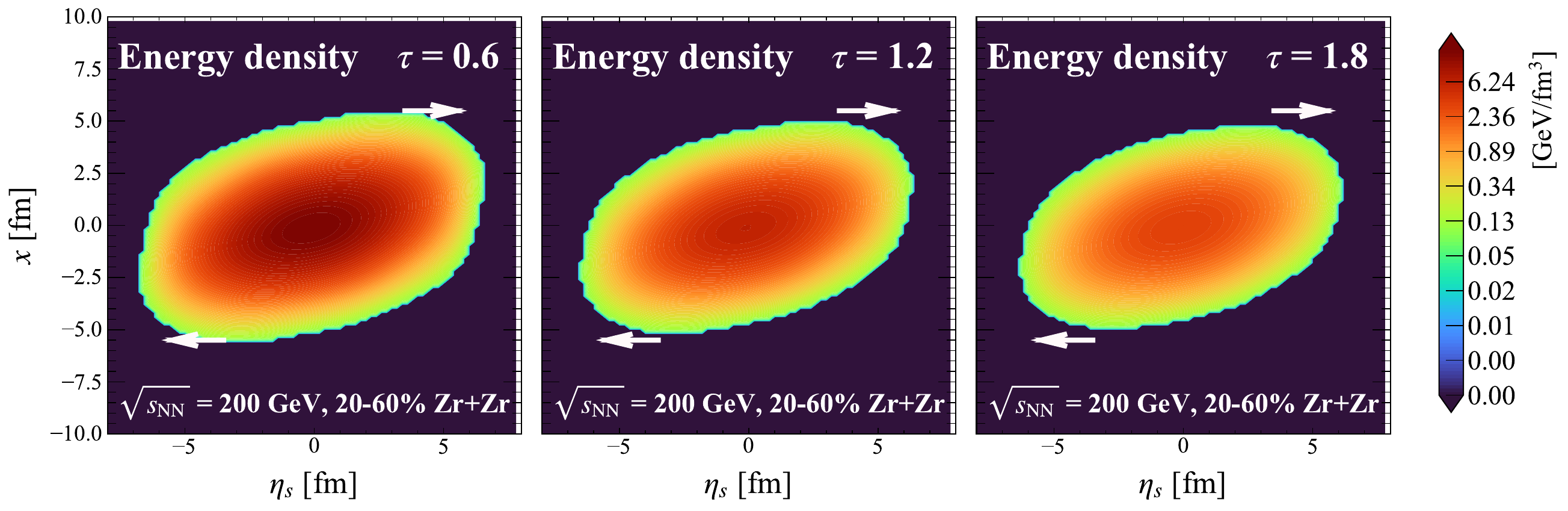} \\
\includegraphics[width=0.85\linewidth]{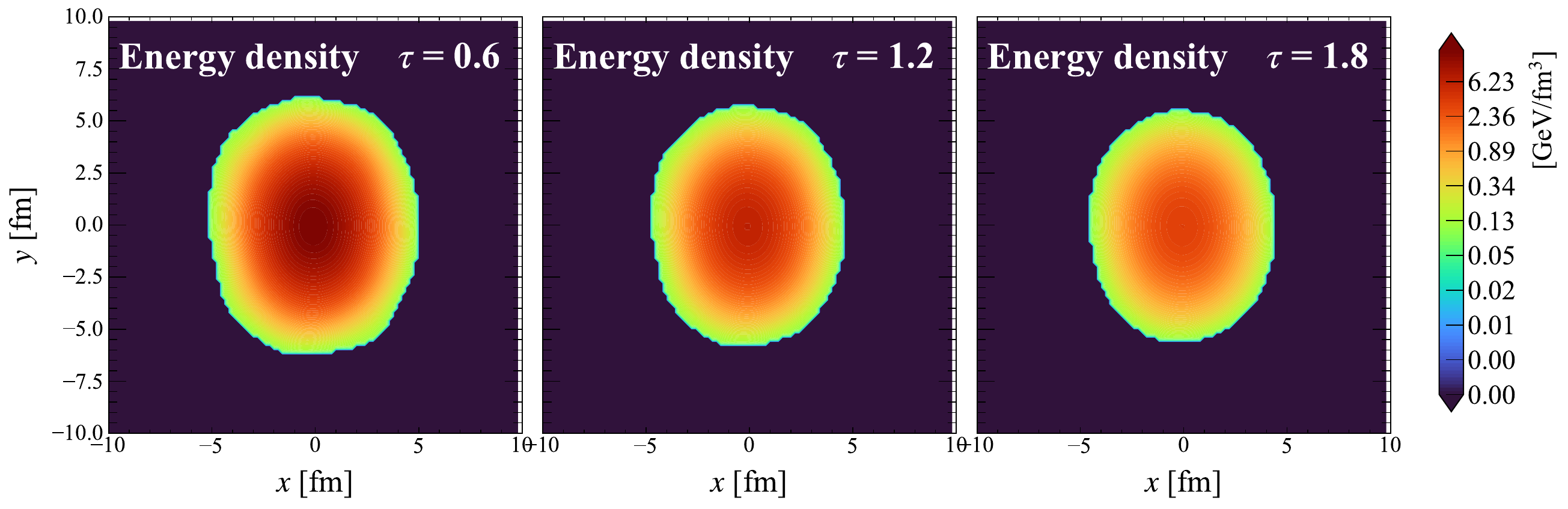}
\end{center}
\caption{(Color online) 
Upper panel: the energy density distribution of the QGP in the reaction plane
($\eta_\text{s}$--$x$ plane, $y=0$) at proper times $\tau = 0.6$, 1.2, and 1.8~fm, for
20--60\% Zr+Zr collisions at $\sqrt{s_\text{NN}}=200$~GeV from the
\trento\ initialization averaged over 1000 events. The counterclockwise tilt of the
fireball with respect to the longitudinal direction, originating from the
$k_\text{T}$-dependent fragmentation regions, is clearly visible and gradually dilutes
as the system expands. Lower panel: the corresponding energy density distributions in
the transverse ($x$--$y$) plane at $\eta_\text{s}=0$, illustrating the radial expansion
of the almond-shaped overlap region. The color scale spans from 0 to 6.24~GeV/fm$^3$.}
\label{f:ed}
\end{figure*}

Building on the participant thickness formalism of \trento, the three-dimensional
energy density profile at proper time $\tau_0$ is constructed as
\begin{equation}
\begin{aligned}
\varepsilon_{\text{IC}}(\vec{x}_\perp,\eta_s) &= \underbrace{N_{\text{fb}}\sqrt{T_A(\vec{x}_\perp)T_B(\vec{x}_\perp)}f_{\text{fb}}({\eta_s - \eta_{cm}(\vec{x}_\perp)} )}_{\text{Central fireball}} \\ 
&+ \sum_{X=A,B}\underbrace{\frac{k_\text{T}}{2N_{\text{frag}}}F_X(\vec{x}_\perp)f_{\text{frag}}^X({e^{-\eta_{max}\pm\eta_s}})}_{\text{Fragmentation regions}}.    
\end{aligned}
\label{eq:tot_energy}
\end{equation}
At any transverse position $\vec{x}_\perp$, \trento\ models the longitudinal structure
as a Gaussian-like central fireball around midrapidity, flanked by two fragmentation
regions inspired by the limiting fragmentation hypothesis. The $\pm$ signs distinguish
the forward ($+$) region (nucleus $A$) from the backward ($-$) region (nucleus $B$).
The nuclear thickness functions $T_X$ incorporate sub-nucleonic structure through $n_c$
constituent partons with Gaussian smearing:
\begin{equation}
T_X(\vec{x}_\perp) = \sum_{p\in X}\frac{1}{n_c}\sum_{c\in p}\gamma_c\;
\frac{\exp\bigl(-|\vec{x}_\perp - \vec{x}_p - \vec{s}_c|^2/(2v^2)\bigr)}{2\pi v^2},
\end{equation}
where the index $p$ runs over all participant nucleons in nucleus $X$, and the index
$c$ runs over the $n_c$ constituents within each nucleon $p$. The nucleon positions
$\vec{x}_p$ are sampled from the deformed Woods-Saxon distribution of
Eq.~(\ref{eq:WS_deformed}). The transverse positions of the constituents, $\vec{s}_c$,
are sampled from a normal distribution $\mathcal{N}(0, \sigma_s^2)$, with width
$\sigma_s = \sqrt{(w^2 - v^2)/(1-1/n_c)}$. The constituent width $v$ is determined by
the structure parameter $\chi$ from Table~\ref{tab:params3d} via $v = \chi w
n_c^{1/4}$, ensuring $v < w$. The fraction factor $\gamma_c$ for energy deposition
follows a beta distribution:
\begin{equation}
\gamma_{c}\sim \text{Beta} \bigl\{ \langle\gamma_{c}\rangle (1-k)/k,\; (1-\langle\gamma_{c}\rangle)(1-k)/k \bigr\},
\end{equation}
where the fluctuation parameter $k$ controls the shape of this distribution: large $k$
values skew $\gamma_c$ towards 0 or 1, making an even energy split unlikely, while
small $k$ values favor values near the mean. The mean fraction $\langle\gamma_c\rangle$
is determined self-consistently by the energy conservation condition Eq.~(\ref{eq9}).

The fireball component's rapidity profile combines a deformed Gaussian with
energy-dependent tapering:
\begin{equation}
f_{\text{fb}}(\eta_s) =
\exp\!\left[-\frac{(\eta_s^{2})^{f}}{2\Delta\eta^{2}}\right]\left[1-\Bigl(\frac{\eta_s}{\eta_{\text{max}}}\Bigr)^{4}\,\right]^{4},
\end{equation}
where $\Delta\eta = \eta_{\text{max}} - \nu$ controls the Gaussian width, with the
narrowing parameter $\nu$ fixed to 3 in this work, following
Ref.~\cite{Soeder:2023vdn}. The local center-of-mass rapidity
$\eta_{\text{cm}}(\vec{x}_\perp)$ ensures local momentum conservation:
\begin{equation}
\eta_{\text{cm}}(\vec{x}_\perp) = \text{arctanh}\!\left[\sqrt{1-\frac{4m_p^2}{s_{\text{NN}}}}\;
\frac{T_A(\vec{x}_\perp) - T_B(\vec{x}_\perp)}{T_A(\vec{x}_\perp) + T_B(\vec{x}_\perp)}\right].
\end{equation}
For isobaric Zr+Zr collisions, $T_A$ and $T_B$ are symmetric on average, yet
event-by-event fluctuations in the sub-nucleonic structure and the deformed nuclear
shape can induce local imbalances that generate finite $\eta_{\text{cm}}$ and
contribute to the tilted geometry of the fireball.

The fragmentation regions in Eq.~(\ref{eq:tot_energy}) are described using modified
parton distribution functions scaled by the transverse momentum scale $k_\text{T}$:
\begin{equation}
f_{\text{frag}}^X(x) = (-\ln x)^{\alpha}\,x^{\beta+1}\,
\exp\!\left(-\frac{2k_\text{T}}{x\sqrt{s_{\text{NN}}}}\right),
\end{equation}
where the momentum fraction $x$ depends on spacetime rapidity $\eta_s$ via $x =
e^{-\eta_{\text{max}} \pm \eta_s}$. The thickness functions for the $X$-going fragments
are given by:
\begin{equation}
F_X(\vec{x}_\perp) = \sum_{p\in X}\frac{1}{n_c}\sum_{c\in p}(1-\gamma_c)\;
\frac{\exp\bigl(-|\vec{x}_\perp - \vec{x}_p - \vec{s}_c|^2/(2v^2)\bigr)}{2\pi v^2}.
\end{equation}
The normalization factor $N_{\text{frag}}$ in Eq.~(\ref{eq:tot_energy}) is computed by
integrating the fragmentation profile function: $N_{\text{frag}} = \int_{x_0}^1
dx\,f_{\text{frag}}(x)$, where $x_0 = \exp(-\eta_{\text{max}})$~\cite{Soeder:2023vdn}.
The dynamic rapidity window $\eta_{\text{max}}$ evolves with collision energy as
\begin{equation}
\eta_{\text{max}} = \text{arccosh}\!\left(\frac{\sqrt{s_\text{NN}}}{2k_\text{T}}\right).
\label{eq:eta_max}
\end{equation}
The global energy-momentum conservation in the model is ensured by
\begin{equation}
\langle{\gamma_c}\rangle\sqrt{s_{\text{NN}}} = N_{\text{fb}} \int_{-\eta_{\text{max}}}^{\eta_{\text{max}}} \cosh\eta_s\,f_{\text{fb}}(\eta_s)\,d\eta_s,
\label{eq9}
\end{equation}
which captures the transition from net-baryon accumulation at RHIC energies to
midrapidity-dominated deposition at LHC energies. The \trento\ initialization
parameters adopted in this study follow the Bayesian calibration of
Ref.~\cite{Soeder:2023vdn} and are listed in Table~\ref{tab:params3d}.

\begin{table}[!th]
\caption{Parameters of the \trento\ initialization and the longitudinal flow extension
         used in this study. The first nine parameters are taken from the Bayesian
         calibration of Ref.~\cite{Soeder:2023vdn}; the last parameter $f_v$ is
         introduced in this work to extend the initialization beyond the Bjorken
         approximation.}
\label{tab:params3d}
\centering\footnotesize
\begin{tabular}{@{}llp{5.2cm}@{}}
\toprule
\hline
Parameter & Value & Influence \\
\hline
\hline
$n_c$ & 16.4 & Tunes initial sub-nucleonic degrees of freedom \\
$w$ [fm] & 1.3 & Sets initial source size $\propto \sqrt{\langle r^2 \rangle}$ \\
$\chi = v/(w n_c^{1/4})$ & 0.5 & Tunes sub-nucleon correlation length \\
$f$ & 1.0 & Controls the central fireball profile \\
$N_{\text{fb}}$ & 9.6 & Central fireball energy scale  \\
$k$ & 0.104 & Controls fireball/fragment energy fluctuation \\
$k_\text{T}$ [GeV] & 0.33 & Determines $\eta_{\text{max}}$ and tilted geometry \\
$\alpha$ & 4.6 & Controls the shape of the fragmentation profile \\
$\beta$ & 0.19 & Controls the shape of the fragmentation profile \\
$f_v$ & 0.10 & Fraction of $\eta_{\text{cm}}$ deposited into longitudinal flow \\
\hline
\hline
\bottomrule
\end{tabular}
\end{table}

To illustrate the three-dimensional structure of the initial condition, we present in
Fig.~\ref{f:ed} the smooth energy density distributions (averaged 1000 events) of the QGP fireball in 20--60\% Zr+Zr
collisions at $\snn=200$~GeV. The upper panel displays the energy density in the
$\eta_\text{s}$--$x$ plane (reaction plane, $y=0$) at three successive proper times
$\tau = 0.6$, 1.2, and 1.8~fm, while the lower panel shows the corresponding
distributions in the $x$--$y$ plane (transverse plane, $\eta_\text{s}=0$). From the
upper panel, one can clearly observe a tilted geometry of the fireball with respect to
the longitudinal direction, which originates from the $k_\text{T}$-dependent
fragmentation regions in the \trento\ framework. As the system expands
hydrodynamically, this tilt gradually dilutes but remains visible throughout the early
evolution. From the lower panel, the characteristic almond shape of the overlap region
in non-central collisions is visible, and its radial expansion from $\tau = 0.6$ to
1.8~fm reflects the buildup of transverse collective flow driven by the pressure
gradient of the QGP. Both the tilted longitudinal geometry and the anisotropic
transverse profile can contribute to the vorticity field that polarizes hyperons.

The longitudinal structure of the QGP fireball in the \trento\ initialization is
characterized by central and fragmentation regions. The $k_T$ parameter controls the
fireball tilt through the fragmentation profile. However, the \trento\ model in its
original formulation lacks an initial longitudinal flow velocity gradient, which is
known to crucially affect the vorticity field and hyperon
polarization~\cite{Li:2022pyw,Ryu:2021lnx,Alzhrani:2022dpi}. We therefore introduce the
longitudinal flow gradient into the initialization by extending beyond the Bjorken flow
approximation. Following the approach developed in
Refs.~\cite{Shen:2020jwv,Jiang:2021foj,Jiang:2023vxp}, we modify the energy-momentum
tensor components at the initial proper time $\tau_0$:
\begin{align}
\label{eq:Ttautau}
T^{\tau\tau} &= \varepsilon_{\text{IC}}(\vec{x}_\perp,\eta_s) \cosh(y_\text{L}), \\
\label{eq:Ttaueta}
T^{\tau\eta_s} &= \frac{1}{\tau_{0}} \varepsilon_{\text{IC}}(\vec{x}_\perp,\eta_s) \sinh(y_\text{L}),
\end{align}
where the rapidity variable $y_\text{L}$ is parameterized as
\begin{equation}
y_\text{L} = f_{v} \cdot \eta_{\text{cm}}(\vec{x}_\perp).
\label{eq:yl}
\end{equation}
Here, $\eta_{\text{cm}}(\vec{x}_\perp)$ is the local center-of-mass rapidity already
provided by the \trento\ model, which encodes the imbalance between the participant
thickness functions of the two colliding nuclei. The parameter $f_v \in [0, 1]$
controls the fraction of this local rapidity that is deposited into the initial
longitudinal flow velocity of the medium. With Eqs.~(\ref{eq:Ttautau})
and~(\ref{eq:Ttaueta}), the initial fluid velocity in the $\eta_s$ direction is given
by
\begin{equation}
v_{\eta_s} = \frac{T^{\tau\eta_s}}{T^{\tau\tau} + P},
\end{equation}
where $P$ is the local pressure obtained from the equation of state. The initial fluid
velocity in the transverse plane is set to zero via $T^{\tau x} = T^{\tau y} = 0$, as
transverse pre-flow has negligible impact on the polarization observables studied in
this work.

Physically, the parameter $f_v$ governs the magnitude of the longitudinal velocity
gradient $\partial v_{\eta_s}/\partial x$ in the initial state. A larger $f_v$ enhances
the local vorticity of the QGP along the out-of-plane ($-\hat{y}$) direction through
$\omega_{\alpha\beta} \propto \partial_{[\alpha}u_{\beta]}$, which directly increases
the hyperon polarization via the spin-vorticity coupling~\cite{Becattini:2013fla}. When
$f_v = 0$, one recovers the Bjorken flow scenario with
$y_\text{L}=0$~\cite{Shen:2020jwv}. In this work, we adopt $f_v = 0.10$, which is
constrained by comparing our calculation of the $\Lambda$ global polarization with
experimental data, as will be discussed in Sec.~\ref{section3}. Using the modified
energy-momentum tensor components in Eqs.~(\ref{eq:Ttautau})--(\ref{eq:Ttaueta}), the
local energy density $\varepsilon_0$ fed into the hydrodynamic evolution is
recalculated from $T^{\tau\tau}$ and $T^{\tau\eta_s}$ via the root-finding algorithm
described in Ref.~\cite{Ryu:2021lnx}. This self-consistently incorporates the
longitudinal flow gradient into the initial condition while preserving the overall
energy-momentum conservation of the \trento\ framework.

\subsection{CLVisc hydrodynamic simulations}

Starting with the \trento\ initial condition constructed in the previous subsection, we
utilize a (3+1)-D viscous hydrodynamic model CLVisc to describe the evolution of the
QGP medium and its subsequent particlization. The CLVisc model is described in full
detail in Refs.~\cite{Pang:2018zzo,Wu:2021fjf}; here we summarize only its main
features. The CLVisc code solves the local energy-momentum and net baryon conservation
equations
\begin{align}
\nabla_{\mu} T^{\mu\nu} &= 0, \\
\nabla_{\mu} J^{\mu} &= 0,
\end{align}
where the energy-momentum tensor $T^{\mu\nu}$ and the net baryon current $J^{\mu}$ are
defined as
\begin{align}
T^{\mu\nu} &= \varepsilon u^{\mu}u^{\nu} - (P + \Pi)\Delta^{\mu\nu} + \pi^{\mu\nu}, \\	
J^{\mu} &= n u^{\mu} + V^{\mu},
\end{align}
with $\varepsilon$, $P$, $\Pi$, $n$, $u^{\mu}$, $\pi^{\mu\nu}$, $V^{\mu}$ being the
local energy density, equilibrium pressure, bulk viscous pressure, net baryon density,
flow velocity field, shear stress tensor, and baryon diffusion current, respectively.
The projection tensor is given by $\Delta^{\mu\nu} = g^{\mu\nu} - u^{\mu}u^{\nu}$ with
the metric tensor $g^{\mu\nu} = \text{diag}(1,-1,-1,-1)$. At $\snn = 200$~GeV the net
baryon density at midrapidity is negligible, and we set $n = 0$ and $V^{\mu}=0$
throughout the hydrodynamic evolution.

The dissipative currents $\pi^{\mu\nu}$ and $\Pi$ are evolved according to the
Israel-Stewart-like second-order hydrodynamic equations~\cite{Denicol:2018wdp}. The
specific shear viscosity is fixed as $\eta/s = 0.08$, consistent with Bayesian
extractions from soft hadron observables at RHIC
energies~\cite{Ryu:2017qzn,Bernhard:2019bmu,JETSCAPE:2020shq}. For the specific bulk
viscosity, we explore three constant values together with a temperature-dependent
parametrization. Defining $x = T/T_0$ with $T_0 = 0.18$~GeV, and the temperature
boundaries $T_a = 0.995\,T_0$, $T_b = 1.05\,T_0$, the temperature-dependent form reads
\begin{equation}
(\zeta/s)(T) =
\begin{cases}
\begin{aligned}
0.03 &+ 0.9\,e^{(x-1)/0.0025} \\
     &+ 0.22\,e^{(x-1)/0.022}
\end{aligned}
& T < T_a, \\[3ex]
-13.45 + 27.55\,x - 13.77\,x^2, & T_a \le T \le T_b, \\[3ex]
\begin{aligned}
0.001 &+ 0.9\,e^{-(x-1)/0.025} \\
      &+ 0.25\,e^{-(x-1)/0.13}
\end{aligned}
& T > T_b.
\end{cases}
\label{eq:zetas}
\end{equation}
This temperature-dependent parametrization follows the Duke/Bayesian form used in
Refs.~\cite{Bernhard:2016tnd,Ryu:2017qzn}; it peaks near the QCD crossover temperature
$T_0 \approx 180$~MeV and falls off exponentially on either side. The relaxation time
is taken as $\tau_\Pi = \zeta/[15(\varepsilon+P)(1/3 - c_s^2)^2]$ with $c_s$ the speed
of sound from the EOS~\cite{Denicol:2018wdp}.

The hydrodynamic equations are closed by the HotQCD-2014 equation of
state~\cite{Monnai:2019hkn,Monnai:2021kgu}. The freeze-out hypersurface is determined
by a constant energy density $e_{\text{frz}} = 0.4$~GeV/fm$^3$, which corresponds to a
hadronization temperature of approximately $T_H \simeq 155$~MeV in the present EOS
setup and varies only weakly across the sampled surface. This is the sense in which we
treat the freeze-out hypersurface as approximately isothermal in the polarization
calculation. On this hypersurface, hadrons are sampled using the Cooper-Frye formalism
with out-of-equilibrium corrections from both shear ($\delta f_\pi$) and bulk ($\delta
f_\Pi$) viscosities~\cite{Pang:2018zzo,Wu:2021fjf}. Resonance decays are included for
the bulk hadron spectra following Ref.~\cite{Pang:2018zzo}, whereas feed-down effects
on the polarization observables are not incorporated. Hadronic rescattering after
freeze-out is also not incorporated in the present study.

\subsection{Spin polarization in the isothermal equilibrium scenario}

In non-central heavy-ion collisions, the large orbital angular momentum of the QGP
polarizes the constituent quarks through spin-orbit
coupling~\cite{Liang:2004ph,Liang:2004xn}, which is transferred to final-state hyperons
via spin-vorticity coupling~\cite{Becattini:2013fla}. The polarization pseudo-vector
for spin-$1/2$ fermions is given by the modified Cooper-Frye
formula~\cite{Becattini:2013fla,Fang:2016vpj}:
\begin{equation}
\mathcal{S}^{\mu}(\mathbf{p}) = \frac{\int d\Sigma\cdot p\;\mathcal{J}_{5}^{\mu}(p,X)}{2m\,\Phi(\mathbf{p})},
\label{eq:CS}
\end{equation}
where $d\Sigma^{\mu}$ is the normal vector of the freeze-out hypersurface element,
$p^{\mu}=(\sqrt{\mathbf{p}^{2}+m^{2}},\mathbf{p})$ is the four-momentum, $m$ is the
mass of the fermion, and $\Phi(\mathbf{p}) = \int d\Sigma\cdot p\,f_{\rm eq}$ with
$f_{\rm eq} = [\exp(p^{\mu}u_{\mu}/T) + 1]^{-1}$ the Fermi-Dirac distribution. The
baryon chemical potential is negligible at $\snn = 200$~GeV.

Inserting the axial charge current density $\mathcal{J}_{5}^{\mu}$ from quantum kinetic
theory~\cite{Hidaka:2017auj,Yi:2021ryh}, the polarization pseudo-vector decomposes into
thermal vorticity and thermal shear
contributions~\cite{Yi:2021unq,Yi:2021ryh,Wu:2022mkr}:
\begin{eqnarray}
\mathcal{S}^{\mu}(\mathbf{p}) &=& \mathcal{S}_{\text{thermal}}^{\mu}(\mathbf{p}) + \mathcal{S}_{\text{shear}}^{\mu}(\mathbf{p}),
\label{eq:total}
\end{eqnarray}
with
\begin{eqnarray}
\mathcal{S}_{\text{thermal}}^{\mu}(\mathbf{p}) &=& \hbar\int d\Sigma\cdot\mathcal{N}_{p}\;\frac{1}{2}\,\epsilon^{\mu\nu\alpha\beta}p_{\nu}\varpi_{\alpha\beta},
\nonumber\\[4pt]
\mathcal{S}_{\text{shear}}^{\mu}(\mathbf{p}) &=& \hbar\int d\Sigma\cdot\mathcal{N}_{p}\;\frac{\epsilon^{\mu\nu\alpha\beta}p_{\nu}u_{\beta}}{(u\cdot p)}\,p^{\sigma}\xi_{\sigma\alpha},
\label{eq:th_shear}
\end{eqnarray}
where $\mathcal{N}_{p}^{\mu} = p^{\mu}f_{\rm eq}(1-f_{\rm eq})/[4m\Phi(\mathbf{p})]$,
and the tensors
\begin{equation}
\begin{aligned}
\varpi_{\alpha\beta} = \frac{1}{2}\!\left[\partial_{\alpha}\!\left(\frac{u_{\beta}}{T}\right) - \partial_{\beta}\!\left(\frac{u_{\alpha}}{T}\right)\right],\\
\xi_{\alpha\beta} = \frac{1}{2}\!\left[\partial_{\alpha}\!\left(\frac{u_{\beta}}{T}\right) + \partial_{\beta}\!\left(\frac{u_{\alpha}}{T}\right)\right],
\label{eq:varpi_xi}
\end{aligned}
\end{equation}
are the thermal vorticity and thermal shear tensors. Equivalent decompositions follow
from linear response theory~\cite{Liu:2020dxg,Liu:2021uhn,Fu:2021pok,Fu:2022myl} and
quantum statistical models~\cite{Becattini:2021suc,Becattini:2021iol}.

At the high collision energy considered in this work ($\snn = 200$~GeV), the freeze-out
hypersurface can be well approximated by an isothermal one with constant temperature
$T_H$~\cite{Becattini:2021suc,Becattini:2021iol,Palermo:2024tza,Yi:2024kwu}. In this
\emph{iso-thermal equilibrium} limit, all terms proportional to $\nabla T$ vanish, and
the tensors in Eq.~(\ref{eq:varpi_xi}) simplify to
\begin{equation}
\varpi_{\alpha\beta} \;\longrightarrow\; \frac{\omega_{\alpha\beta}}{T_H},\qquad
\xi_{\alpha\beta} \;\longrightarrow\; \frac{\Xi_{\alpha\beta}}{T_H},
\label{eq:iso_limit}
\end{equation}
where
\begin{equation}
\omega_{\alpha\beta} = \frac{1}{2}\bigl(\partial_{\alpha}u_{\beta} - \partial_{\beta}u_{\alpha}\bigr),\qquad
\Xi_{\alpha\beta} = \frac{1}{2}\bigl(\partial_{\alpha}u_{\beta} + \partial_{\beta}u_{\alpha}\bigr),
\label{eq:vortshear}
\end{equation}
are the kinematic vorticity and shear tensors. Substituting Eq.~(\ref{eq:iso_limit})
into Eqs.~(\ref{eq:th_shear}) yields the isothermal polarization formulae employed in
this work:
\begin{eqnarray}
\mathcal{S}_{\text{thermal}}^{\mu}(\mathbf{p})\big|_{\text{iso}} &=& 
\frac{\hbar}{2T_H}\int d\Sigma\cdot\mathcal{N}_{p}\;\epsilon^{\mu\nu\alpha\beta}p_{\nu}\,\omega_{\alpha\beta},
\label{eq:S_thermal_iso}
\\[6pt]
\mathcal{S}_{\text{shear}}^{\mu}(\mathbf{p})\big|_{\text{iso}} &=& 
\frac{\hbar}{T_H}\int d\Sigma\cdot\mathcal{N}_{p}\;\frac{\epsilon^{\mu\nu\alpha\beta}p_{\nu}u_{\beta}}{(u\cdot p)}\,p^{\sigma}\,\Xi_{\sigma\alpha}.
\label{eq:S_shear_iso}
\end{eqnarray}
In our numerical implementation, these two contributions are evaluated separately on
the freeze-out hypersurface using pre-computed tensors $\omega_{\alpha\beta}$ and
$\Xi_{\alpha\beta}$, corresponding to the `iso-th' and `iso-sh' modes in the
polarization analysis code. For comparison, we also compute the standard thermal
vorticity and thermal shear contributions from Eqs.~(\ref{eq:th_shear}) (denoted as
`th' and `sh' modes), which include the full temperature gradients and will be
contrasted with the isothermal results in Sec.~\ref{section3}.

\begin{figure*}[tbp!]
\begin{center}
\includegraphics[width=0.48\linewidth]{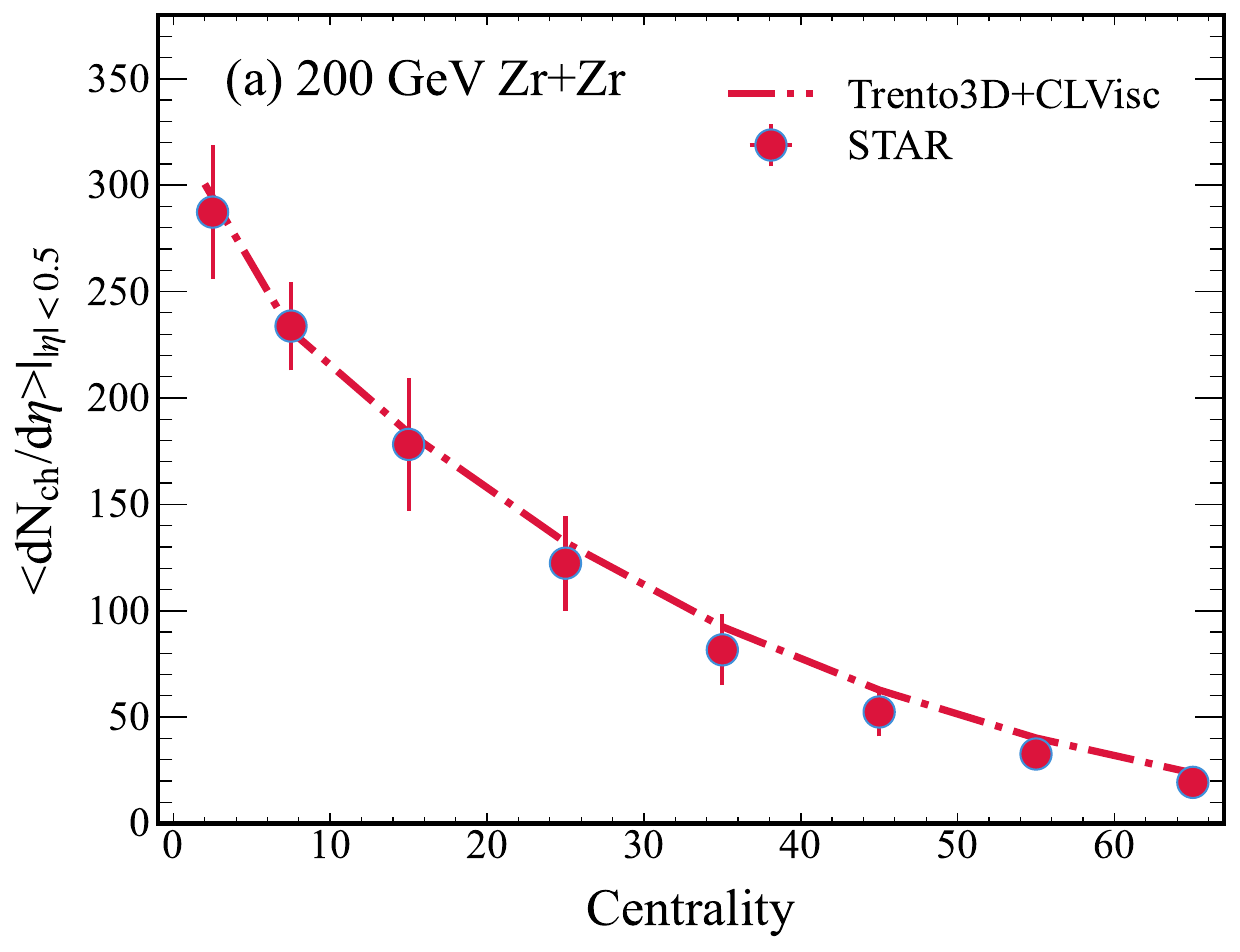}~~
\includegraphics[width=0.48\linewidth]{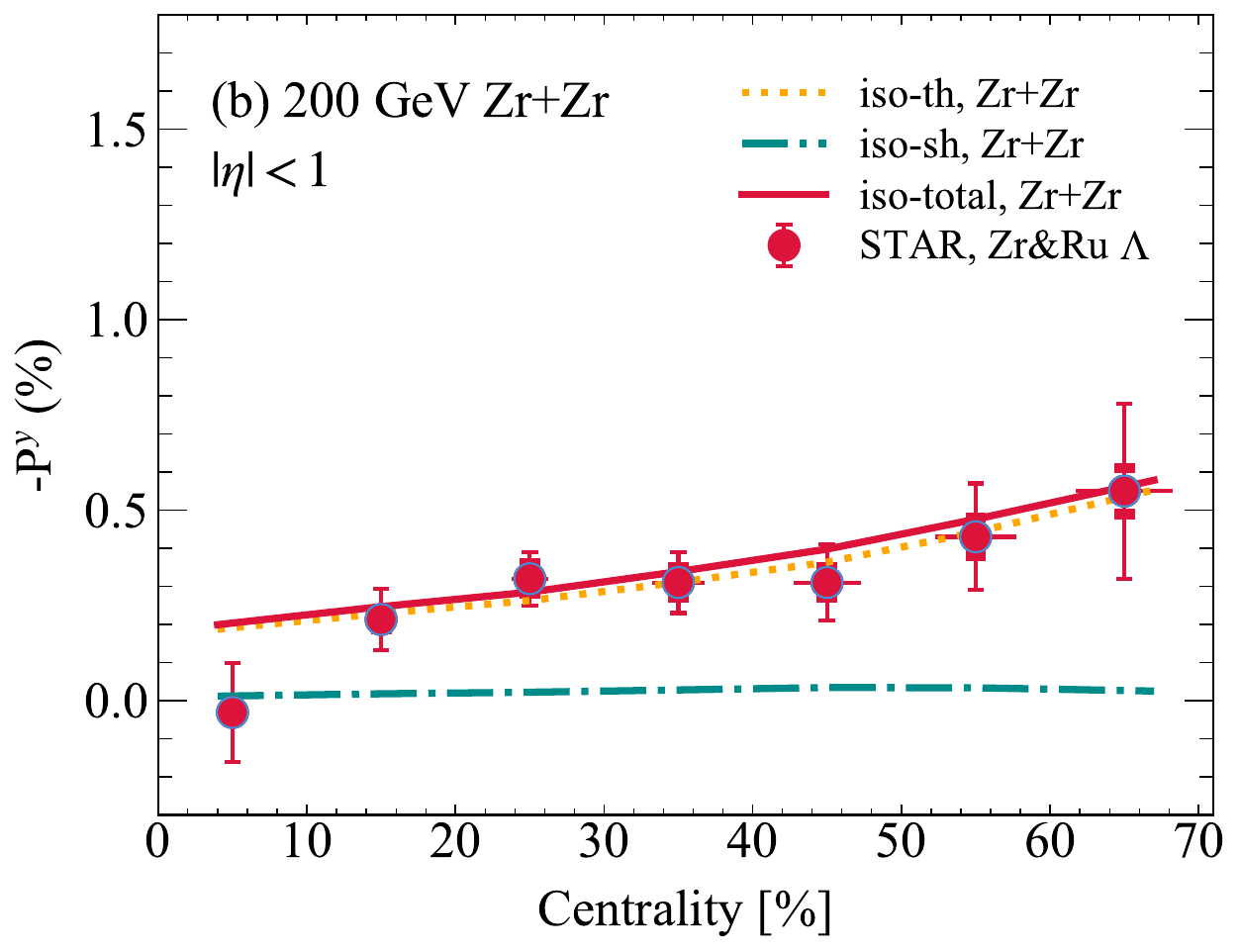} \\
\includegraphics[width=0.48\linewidth]{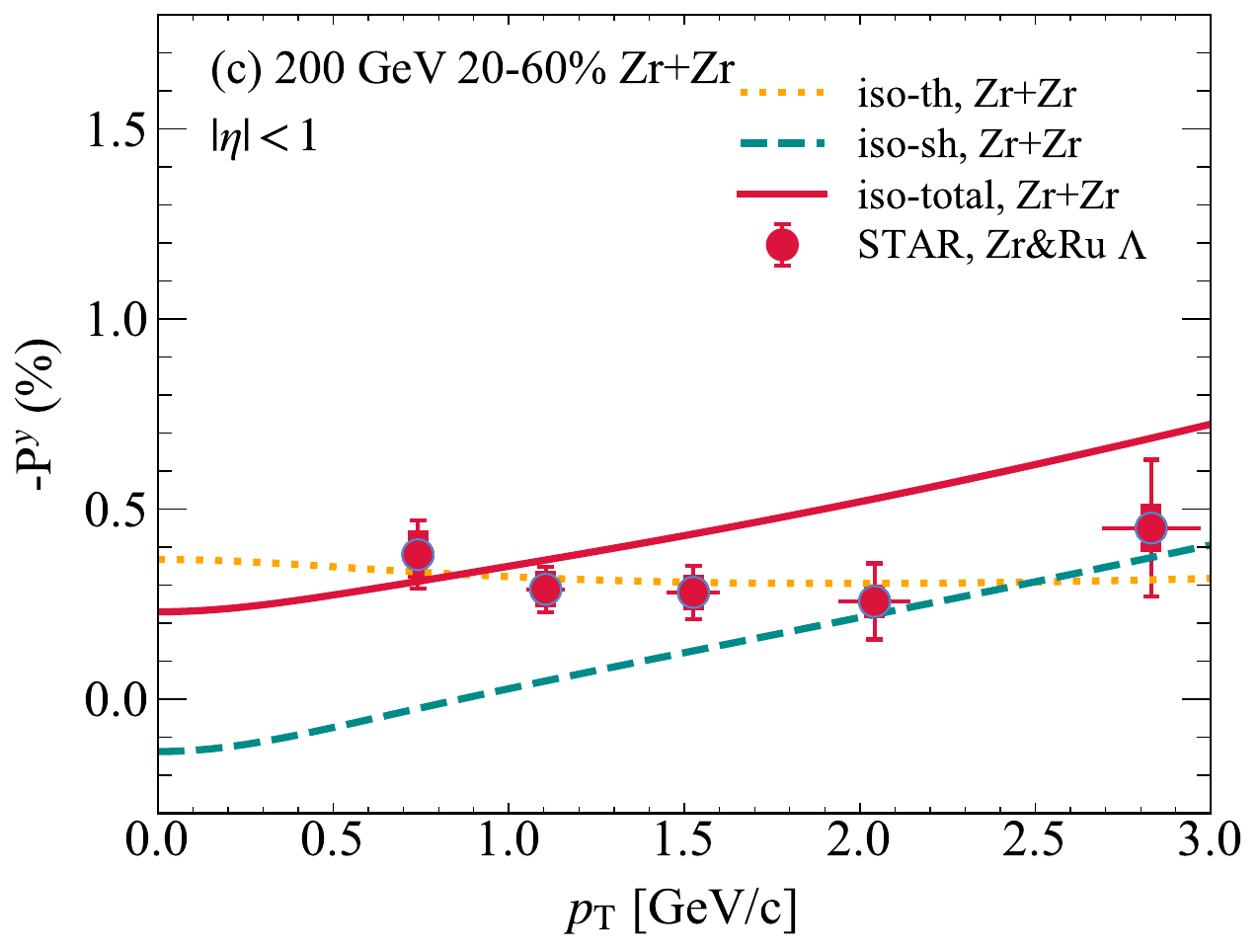}~~
\includegraphics[width=0.48\linewidth]{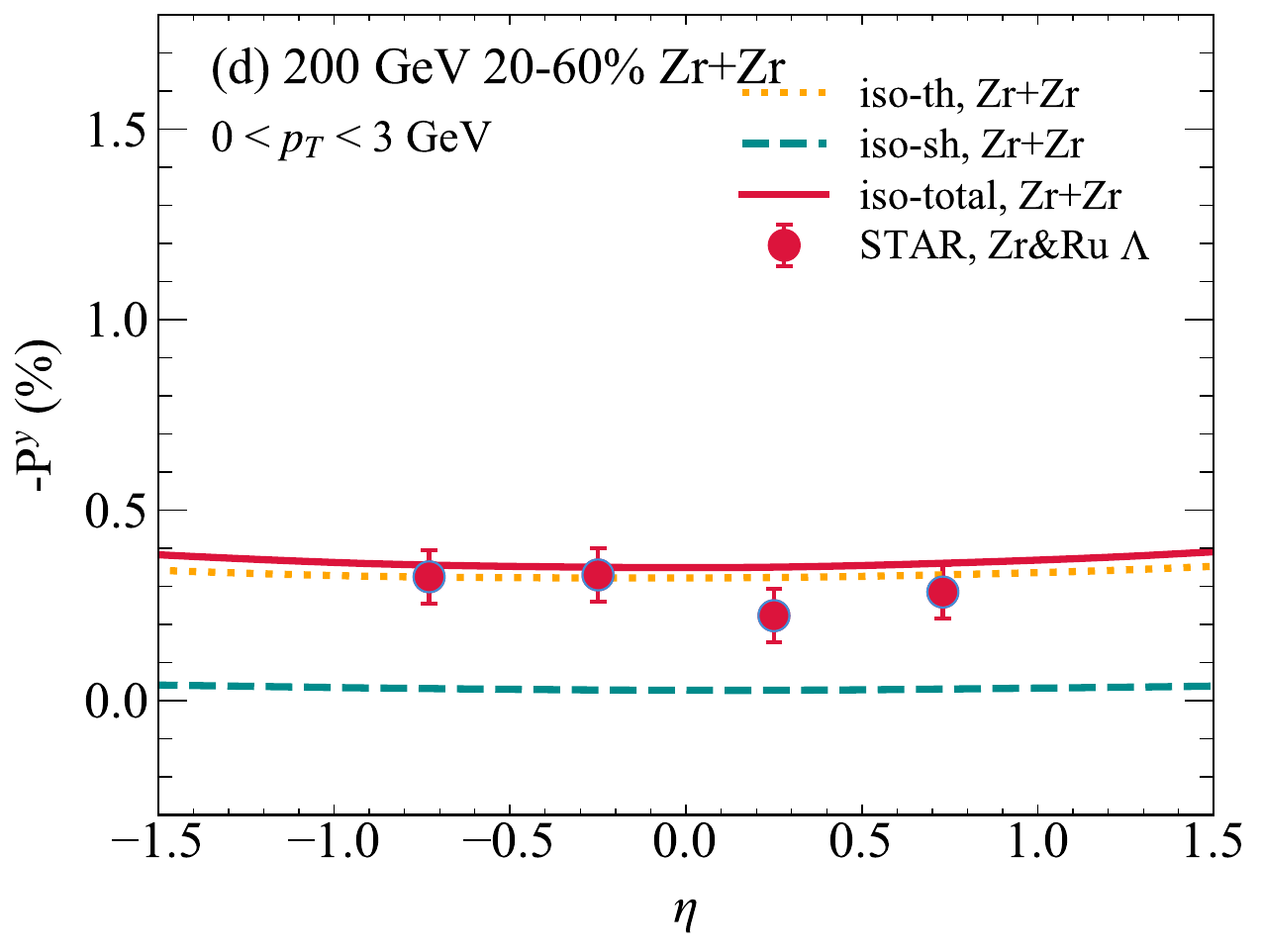}
\end{center}
\caption{(Color online) Baseline observables in Zr+Zr collisions at $\snn=200$~GeV from
         \trento~ + CLVisc compared to STAR data. (a) Charged particle pseudorapidity
         density $\langle dN_{\text{ch}}/d\eta\rangle|_{|\eta|<0.5}$
         vs.\ centrality~\cite{STAR:2021mii}. (b) Global $\Lambda$ polarization
         $-P^{y}$ vs.\ centrality, for $p_T\in[0.5,3.0]$~GeV and
         $|\eta|<1$~\cite{STAR:2025dgs}. (c) $-P^{y}$ vs.\ $p_T$ in 20--60\%
         centrality, $|\eta|<1$. (d) $-P^{y}$ vs.\ $\eta$ in 20--60\% centrality,
         $0<p_T<3$~GeV. In panels (b)--(d), the dotted, dashed, and solid lines
         represent the `iso-th', `iso-sh', and total isothermal polarization,
         respectively. We note that STAR $\Lambda$ data~\cite{STAR:2025dgs} combine
         Zr+Zr and Ru+Ru measurements.}
\label{f:baseline}
\end{figure*}

The polarization vector in the rest frame of $\Lambda$ ($\bar{\Lambda}$) is then
obtained as
\begin{eqnarray}
\vec{P}^{*}(\mathbf{p}) &=& \vec{P}(\mathbf{p}) - \frac{\vec{P}(\mathbf{p})\cdot\vec{p}}{p^{0}(p^{0}+m)}\,\vec{p},
\\[4pt]
P^{\mu}(\mathbf{p}) &\equiv& \frac{1}{s}\,\mathcal{S}^{\mu}(\mathbf{p}),
\end{eqnarray}
with $s=1/2$ the spin of the hyperon and $m=1.116$~GeV its mass. The global
polarization along the direction of the initial orbital angular momentum ($-y$) is
given by
\begin{equation}
\langle -P^{y}\rangle = \frac{\int dy \int p_T dp_T \int d\phi_p\; \Phi(\mathbf{p})\,[-P^{*y}(\mathbf{p})]}
{\int dy \int p_T dp_T \int d\phi_p\; \Phi(\mathbf{p})},
\label{eq:globalP}
\end{equation}
where the negative sign is chosen so that positive values correspond to polarization
along the system's orbital angular momentum. In the present analysis, the kinematic
region for $\Lambda$ and $\bar{\Lambda}$ is $p_T \in [0.5, 3.0]$~GeV and $y \in [-1,
1]$. When comparing with STAR measurements quoted in pseudorapidity windows such as
$|\eta|<1$, we use the standard midrapidity approximation $y \simeq \eta$ for the
measured hyperons in this acceptance. Throughout this paper, $\eta_s$ denotes the
space-time rapidity of the hydrodynamic fields and should not be confused with the
momentum-space variables $y$ and $\eta$. Feed-down contributions to the polarization
observables~\cite{Becattini:2016gvu} and hadronic rescattering after freeze-out are not
included in the current work; for recent discussions of these effects on spin
polarization, we refer to Refs.~\cite{Sung:2024vyc,Palermo:2024tza}.

\section{Numerical results}
\label{section3}
\subsection{Baseline: bulk hadron production and global polarization}
\label{sec:3-1}

We first establish the baseline performance of the \trento\ + CLVisc framework for
Zr+Zr collisions at $\snn=200$~GeV. All polarization results in this section are
obtained in the isothermal scenario (`iso-th', `iso-sh', and `iso-total=iso-th+iso-sh')
with the longitudinal flow parameter $f_v=0.10$ and $k_T=0.33$~GeV, as detailed in
Sec.~\ref{v1section2}.

Figure~\ref{f:baseline} summarizes the baseline comparisons between our model and STAR
measurements for Zr+Zr collisions at $\snn=200$~GeV. Panel (a) shows the charged
particle density at midrapidity as a function of centrality. The \trento\
initialization provides a satisfactory description of the STAR data~\cite{STAR:2021mii}
across the full centrality range, confirming that the overall entropy deposition and
the hydrodynamic expansion of the medium are reliably captured.

Panel (b) presents the centrality dependence of the global $\Lambda$ polarization along
the out-of-plane direction, $-P^{y}$. This provides a quantitative description of
hyperon polarization in isobaric Zr+Zr collisions at RHIC within the \trento\ + CLVisc
framework. The isothermal vorticity contribution (`iso-th', dotted) increases
monotonically from central to peripheral collisions, reflecting the larger orbital
angular momentum deposited into the QGP at larger impact
parameters~\cite{Becattini:2013fla}. The isothermal shear contribution (`iso-sh') is
subdominant: it rises modestly up to 40--50\% centrality and then decreases toward the
most peripheral collisions, where it becomes negligible. Consequently, the total
polarization (solid line) is dominated by the thermal-vorticity term, which is in
agreement with the STAR measurements~\cite{STAR:2025dgs} within their statistical
uncertainties.

Panel (c) shows the $p_T$ dependence of $-P^{y}$ in the 20--60\% centrality bin, where
the interplay between the two polarization sources is clearly exhibited. The thermal
vorticity term decreases steadily with $p_T$. In contrast, the shear term grows
continuously with $p_T$ and overtakes the thermal vorticity contribution near
$p_T\approx 2.5$~GeV. Their combination produces a characteristic total polarization
(solid line) that rises from low to high $p_T$, driven by the growing shear term, and
reaches $\sim 0.7\%$ at $p_T=3$~GeV. This behavior underscores the essential role of
the shear-induced polarization: without it, the calculated $-P^{y}$ would fall
monotonically with $p_T$. The STAR measurements~\cite{STAR:2025dgs}, despite their
limited statistical precision, are consistent with the overall magnitude and the rising
trend predicted by the model.

Panel (d) displays the pseudorapidity dependence of $-P^{y}$ for $0<p_T<3$~GeV in
20--60\% centrality. For the symmetric Zr+Zr system, the polarization is an even
function of $\eta$, and the total polarization (solid) exhibits a shallow midrapidity
dip, remaining nearly flat within $|\eta|<1$. The thermal vorticity and shear
contributions are both nearly flat across $|\eta|<1$, and their superposition naturally
yields this weak concavity. The STAR data~\cite{STAR:2025dgs} are broadly consistent
with this distribution, though the large error bars preclude a detailed discrimination
of the $\eta$ structure. The overall agreement confirms that the initial longitudinal
flow gradient $f_v=0.10$, together with the tilted fireball geometry from the
$k_T$-dependent fragmentation regions, provides a realistic description of the
vorticity field in this isobaric collision system.

In summary, the \trento\ + CLVisc framework with the isothermal polarization scenario
successfully reproduces the charged particle multiplicity and provides a quantitative
description of the global $\Lambda$ polarization in Zr+Zr collisions at $\snn=200$~GeV
across centrality, $p_T$, and pseudorapidity. The $p_T$ dependence, in particular,
reveals the critical role of the shear-induced polarization at intermediate and high
transverse momenta. These baseline results support the key ingredients of the model and
provide a solid foundation for the more detailed studies presented in the following
sections.

\subsection{Sensitivity of global polarization to initial-state parameters}
\label{sec:3-2}

The baseline results presented in Sec.~\ref{sec:3-1} demonstrate that the \trento\ +
CLVisc framework provides a satisfactory description of the global $\Lambda$
polarization in Zr+Zr collisions. In this section, we systematically investigate how
the longitudinal flow fraction $f_v$, the parton transverse momentum scale $k_T$, and
the nuclear structure of the colliding species affect the global polarization. These
three factors influence $-P^{y}$ through different physical mechanisms,
and their imprints on the $p_T$ and $\eta$ dependences offer new insights into
constraining the three-dimensional initial condition of the QGP.

\subsubsection{\texorpdfstring{Dependence on the longitudinal flow fraction
$f_v$}{Dependence on the longitudinal flow fraction fv}}

\begin{figure}[tbp!]
\begin{center}
\includegraphics[width=0.85\linewidth]{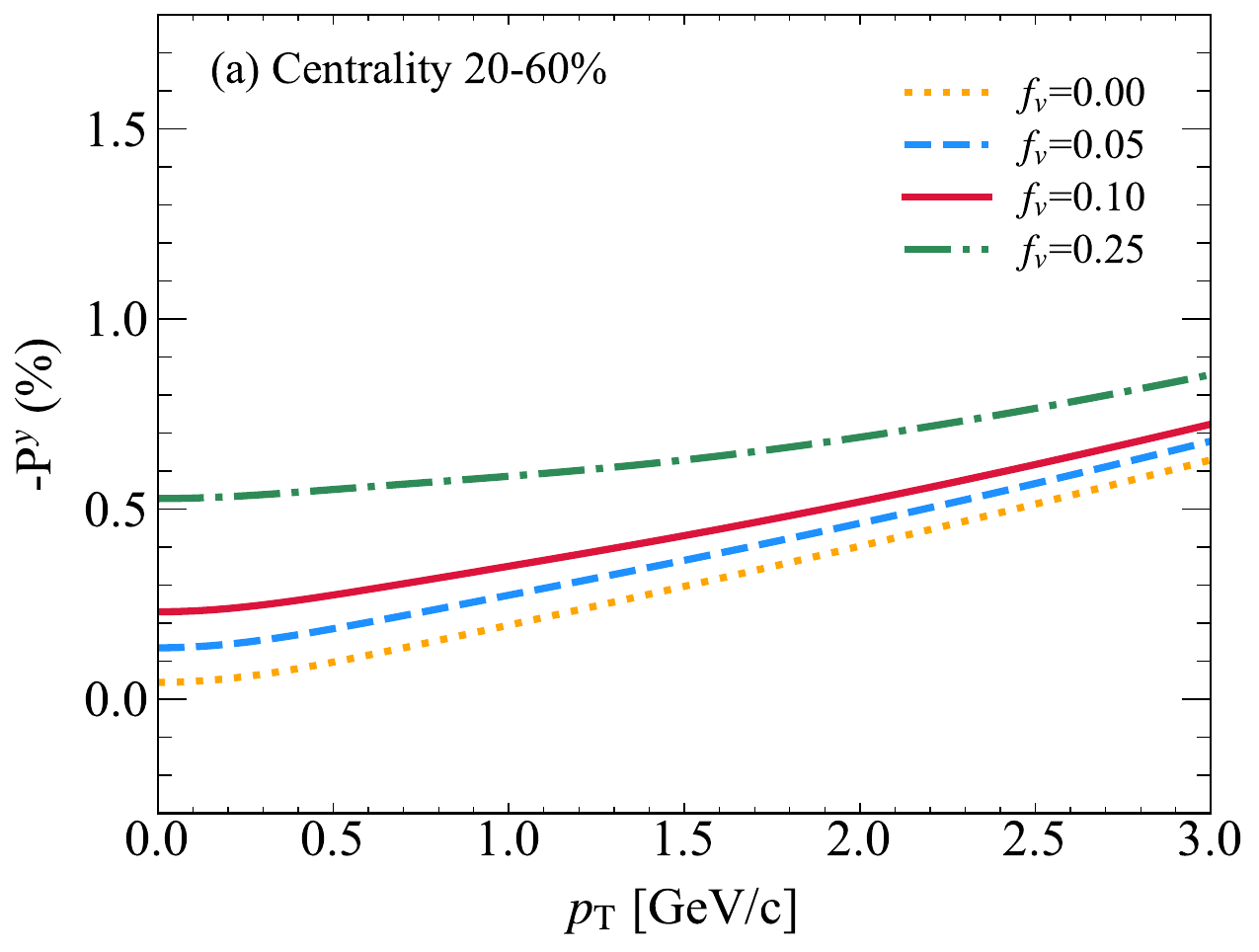} \\
\includegraphics[width=0.85\linewidth]{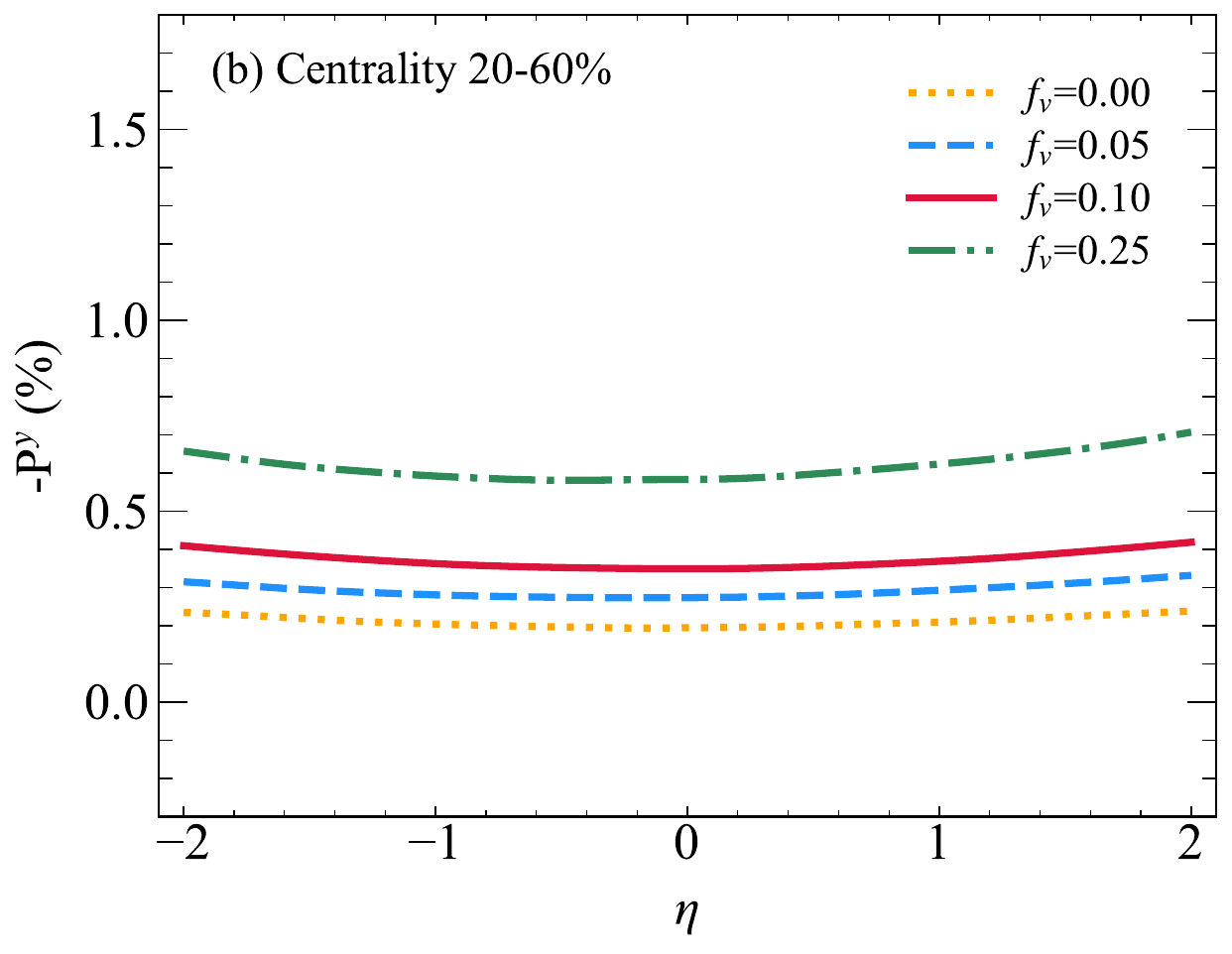}
\end{center}
\caption{(Color online) Dependence of the global $\Lambda$ polarization $-P^{y}$ on the
         longitudinal flow fraction $f_v$ in the isothermal scenario, for 20--60\%
         Zr+Zr collisions at $\snn=200$~GeV. Upper: $-P^{y}$ as a function of $p_T$
         (integrated over $|\eta|<1$). Lower: $-P^{y}$ as a function of $\eta$
         (integrated over $0<p_T<3$~GeV). The four curves correspond to $f_v = 0.00$,
         0.05, 0.10, and 0.25 at fixed $k_T=0.33$~GeV.}
\label{f:fv_scan}
\end{figure}

Figure~\ref{f:fv_scan} displays the $f_v$ dependence of $-P^{y}$ for four values of the
longitudinal flow fraction: $f_v=0.00$, 0.05, 0.10, and 0.25, with $k_T$ fixed at
0.33~GeV. The case $f_v=0$ corresponds to the Bjorken flow limit~\cite{Shen:2020jwv},
where the initial longitudinal velocity gradient is absent and the polarization is
generated solely by the tilted fireball geometry from the $k_T$-dependent fragmentation
regions. Under this condition, the thermal vorticity contribution is small, but the
shear contribution---driven by the geometric tilt of the fireball---remains active,
producing a modest polarization that rises with $p_T$. Nevertheless, this remains below
the experimentally observed magnitude, indicating that a finite longitudinal flow
gradient is required.

Introducing a non-zero $f_v$ enhances the polarization, primarily through the thermal
vorticity channel. At $f_v=0.10$ (our baseline), the $p_T$-integrated $-P^{y}$ at
midrapidity consistent with the STAR data shown in
Fig.~\ref{f:baseline}(b). At $f_v=0.25$, this value grows to $\sim 0.85\%$. This
enhancement is driven by the initial longitudinal velocity gradient $\partial
v_{\eta_s}/\partial x$ introduced through Eqs.~(\ref{eq:Ttautau})--(\ref{eq:yl}), which
strongly amplifies the kinematic vorticity $\omega_{\alpha\beta}$ along the
out-of-plane direction and thereby increases the thermal vorticity contribution. The
shear contribution shows only a mild sensitivity to $f_v$, consistent with the fact
that the shear tensor $\Xi_{\alpha\beta}$ is primarily sourced by the geometric tilt of
the fireball (controlled by $k_T$) rather than by the longitudinal flow gradient. The
dominance of the thermal enhancement at larger $f_v$ has important consequences for the
$p_T$ and $\eta$ dependences of $-P^{y}$, as analyzed below.

Figure~\ref{f:fv_scan}(lower panel) provides further insight into the pseudorapidity
dependence of the $f_v$ effect. $-P^{y}(\eta)$ remains essentially flat across
$|\eta|<1$ for all $f_v$ values, while its overall magnitude increases steadily with
$f_v$. This linear-like
growth is dominated by the thermal vorticity contribution, which is nearly
$\eta$-independent and scales directly with the longitudinal velocity gradient
$\partial v_{\eta_s}/\partial x$. The shear contribution, in contrast, shows only a
weak $\eta$ dependence and a mild sensitivity to $f_v$, consistent with its primary
origin in the geometric tilt of the fireball rather than the flow gradient.

These results establish $f_v$ as a parameter that primarily controls the overall
magnitude of $-P^{y}$ and the relative weight of the thermal-vorticity contribution in
$-P^{y}(p_T)$. In $\eta$, increasing $f_v$ mainly raises the overall level of
$-P^{y}(\eta)$ while keeping the profile nearly flat within $|\eta|<1$. The value
$f_v=0.10$ adopted in this work provides the best overall agreement with the measured
polarization across $p_T$ and $\eta$, and will be used throughout the remainder of this
paper.

\subsubsection{\texorpdfstring{Dependence on the transverse momentum scale
$k_T$}{Dependence on the transverse momentum scale kT}}

\begin{figure}[tbp!]
\begin{center}
\includegraphics[width=0.85\linewidth]{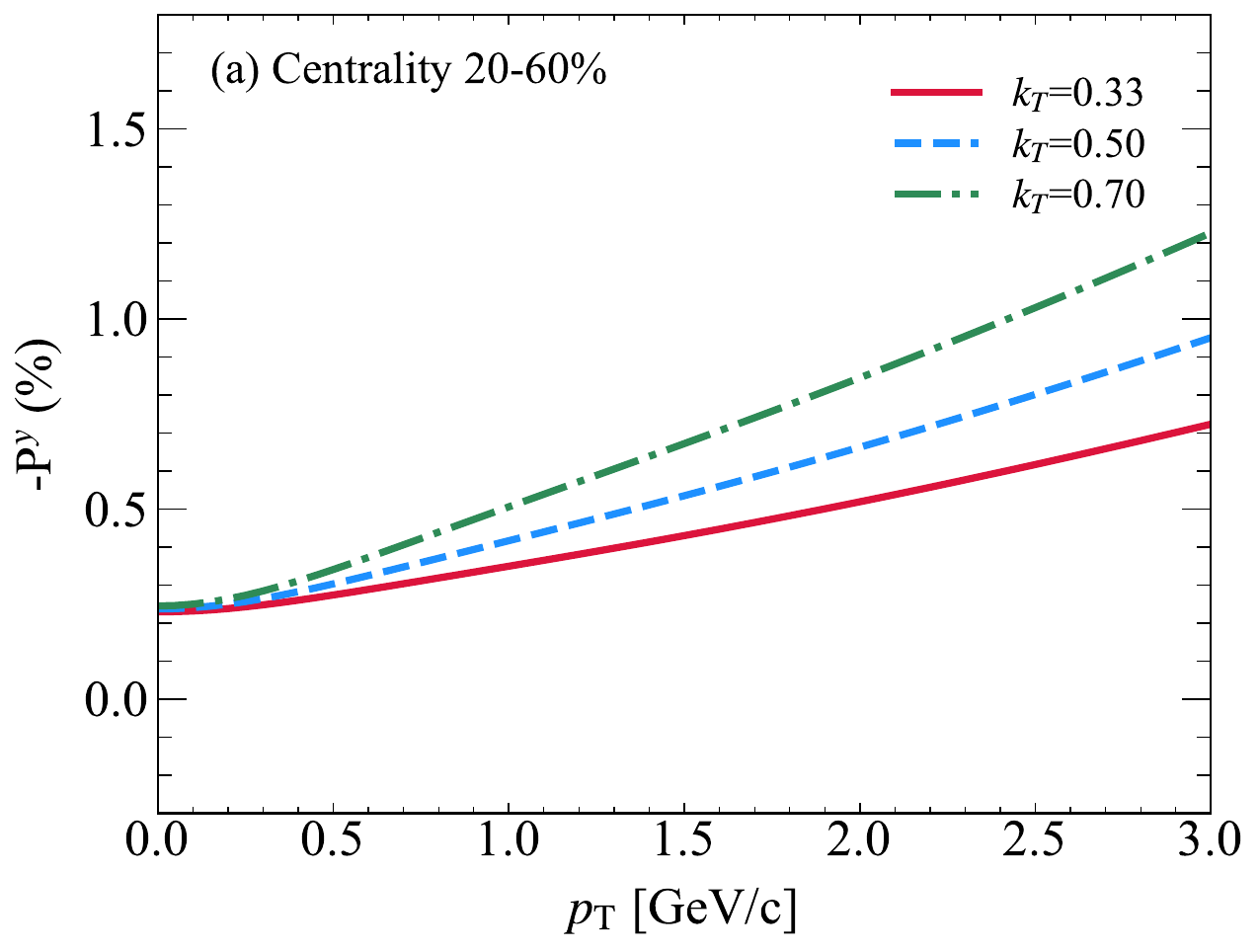} \\
\includegraphics[width=0.85\linewidth]{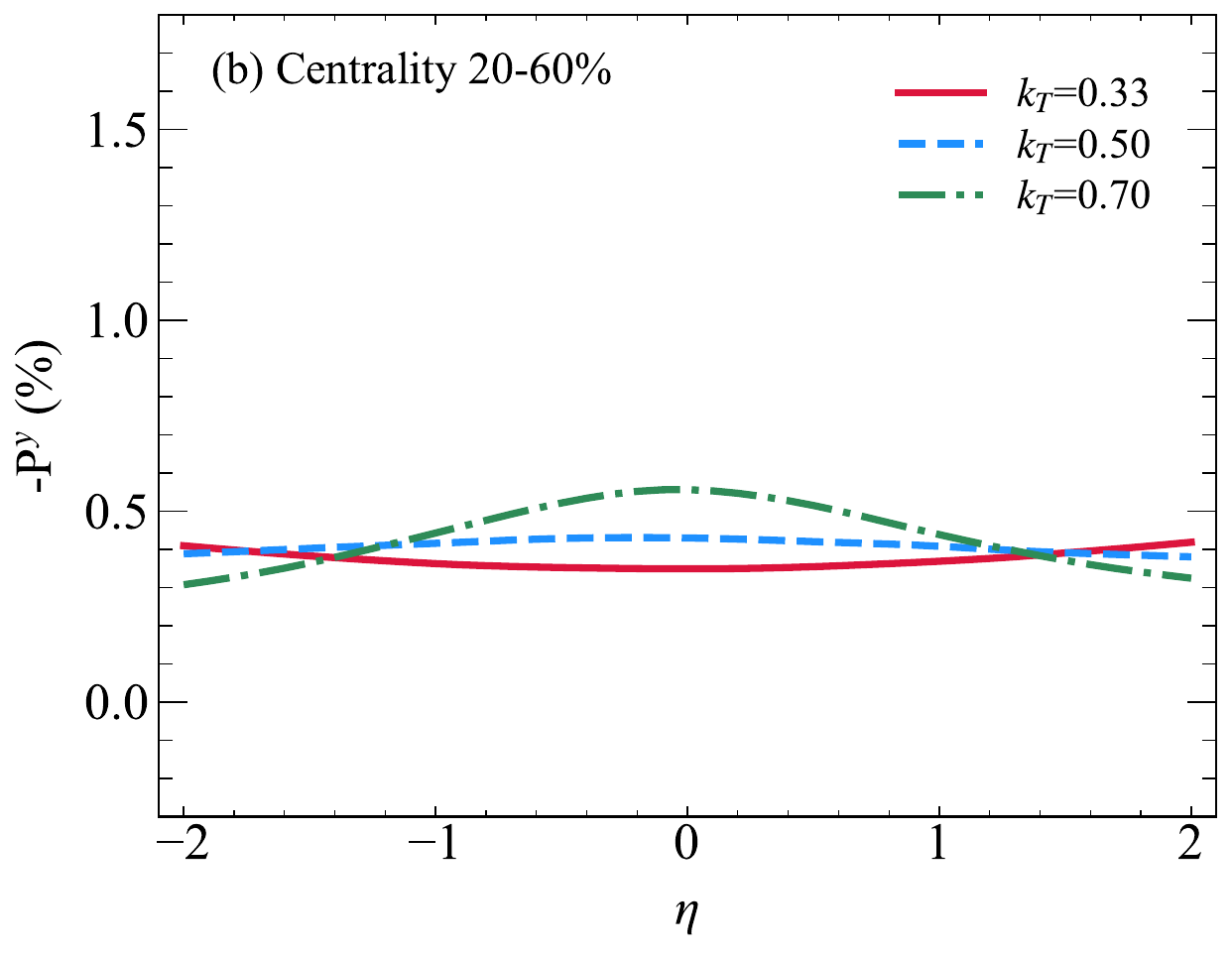}
\end{center}
\caption{(Color online) Dependence of the global $\Lambda$ polarization $-P^{y}$ on the
         parton transverse momentum scale $k_T$ in the isothermal scenario, for
         20--60\% Zr+Zr collisions at $\snn=200$~GeV. Upper: $-P^{y}$ as a function of
         $p_T$ (integrated over $|\eta|<1$). Lower: $-P^{y}$ as a function of $\eta$
         (integrated over $0<p_T<3$~GeV). The three curves correspond to $k_T = 0.33$,
         0.50, and 0.70~GeV at fixed $f_v=0.10$.}
\label{f:kt_scan}
\end{figure}

Figure~\ref{f:kt_scan} presents the $k_T$ dependence of $-P^{y}$ for three values: $k_T
= 0.33$, 0.50, and 0.70~GeV, at fixed $f_v=0.10$. The parameter $k_T$ governs the
longitudinal extent of the fragmentation regions in the \trento\ model [see
Eqs.~(\ref{eq:tot_energy})--(\ref{eq:eta_max})]: a larger $k_T$ reduces the dynamic
rapidity window $\eta_{\text{max}}$ and shifts the fragmentation deposition toward
midrapidity, thereby strengthening the effective tilt relevant for the polarization
signal in the $\eta_s$--$x$ plane.

The impact of $k_T$ on $-P^{y}(p_T)$ is markedly different from that of $f_v$.
Increasing $k_T$ from 0.33 to 0.70~GeV enhances the polarization across the entire
$p_T$ range, with the strongest effect at low $p_T$ where the more strongly tilted
fireball produces a substantially larger shear contribution. As $k_T$ grows, the $p_T$
dependence becomes closer to a monotonic rise. This behavior reflects the direct
sensitivity of the shear tensor $\Xi_{\alpha\beta}$ to the geometric tilt of the
fireball controlled by $k_T$: a more strongly tilted fireball amplifies the
shear-induced polarization, which dominates the overall magnitude and grows with $p_T$.

The lower panel of Fig.~\ref{f:kt_scan} reveals a clear evolution of the $\eta$ profile
with $k_T$. At $k_T=0.33$~GeV, $-P^{y}(\eta)$ is nearly flat across $|\eta|<1$.
As $k_T$ increases, a convex structure gradually develops, with the polarization
peaking at $\eta\approx 0$. At $k_T=0.70$~GeV, this midrapidity enhancement is clearly
visible. This behavior arises because a larger $k_T$ strengthens the effective tilt of
the fireball in the $\eta_s$--$x$ plane, which enhances both the thermal-vorticity and
shear contributions preferentially near midrapidity where the longitudinal velocity
gradient is largest. The transition from flat to convex $\eta$ profiles with increasing
$k_T$ thus reflects the growing influence of the geometric tilt on the spatial
distribution of the polarization.

The comparison of Figs.~\ref{f:fv_scan} and~\ref{f:kt_scan} reveals that $f_v$ and
$k_T$ provide complementary constraints on the initial condition. Increasing $f_v$
mainly raises the overall polarization through the longitudinal flow gradient, while
increasing $k_T$ enhances the polarization by strengthening the geometric tilt and
hence the shear contribution. The latter effect fills in the low-$p_T$ suppression and
drives the $\eta$ profile toward a convex shape. The values $f_v=0.10$ and
$k_T=0.33$~GeV---the former constrained by the polarization data in this work and the
latter by the Bayesian calibration of Ref.~\cite{Soeder:2023vdn}---are both favored by
the comparison with STAR measurements. A simultaneous fit of these two parameters to
precision polarization data would further tighten the constraints on the
three-dimensional initial geometry and flow field of the QGP.

\subsubsection{Dependence on nuclear structure}
\label{subsubsec-3-3}

To assess the sensitivity of global polarization to nuclear geometry, we compare our
baseline $^{96}$Zr calculation (Case~5) with four alternative nuclear structure
configurations for $^{96}$Ru (Cases~1--4) from Ref.~\cite{Jia:2022qgl}. The nuclear
structure parameters for all five configurations are summarized in
Table~\ref{tab:zr_structure}. The five sets span quadrupole deformations from
$\beta_2=0.06$ to $0.162$ and octupole deformations from $\beta_3=0.00$ to $0.20$,
together with modest variations in the half-radius $R_0$ and surface diffuseness $a_0$.
The Woods-Saxon distribution of Eq.~(\ref{eq:WS_deformed}) with these parameters is
used as input to \trento\ for generating the initial nucleon positions, while all other
model parameters are kept fixed at $f_v=0.10$ and $k_T=0.33$~GeV.

\begin{table}[!h]
\centering
\caption{Nuclear structure parameters for the five configurations used in this
         comparison. Case~5 corresponds to $^{96}$Zr and serves as our baseline;
         Cases~1--4 are alternative nuclear structure configurations for $^{96}$Ru
         taken from Ref.~\cite{Jia:2022qgl}.}
\label{tab:zr_structure}
\begin{tabular}{lcccc}
\toprule
\hline
& $R_{0}$ (fm) & $a_{0}$ (fm) & $\beta_{2}$ & $\beta_{3}$ \\
\midrule
\hline
\hline
Case~1 (Ru1)  & 5.09 & 0.46 & 0.162 & 0 \\
Case~2 (Ru2)  & 5.09 & 0.46 & 0.06  & 0 \\
Case~3 (Ru3)  & 5.09 & 0.46 & 0.06  & 0.20 \\
Case~4 (Ru4)  & 5.09 & 0.52 & 0.06  & 0.20 \\
Case~5 (Zr)  & 5.02 & 0.52 & 0.06  & 0.20 \\
\hline
\hline
\bottomrule
\end{tabular}
\end{table}

\begin{figure}[tbp!]
\begin{center}
\includegraphics[width=0.85\linewidth]{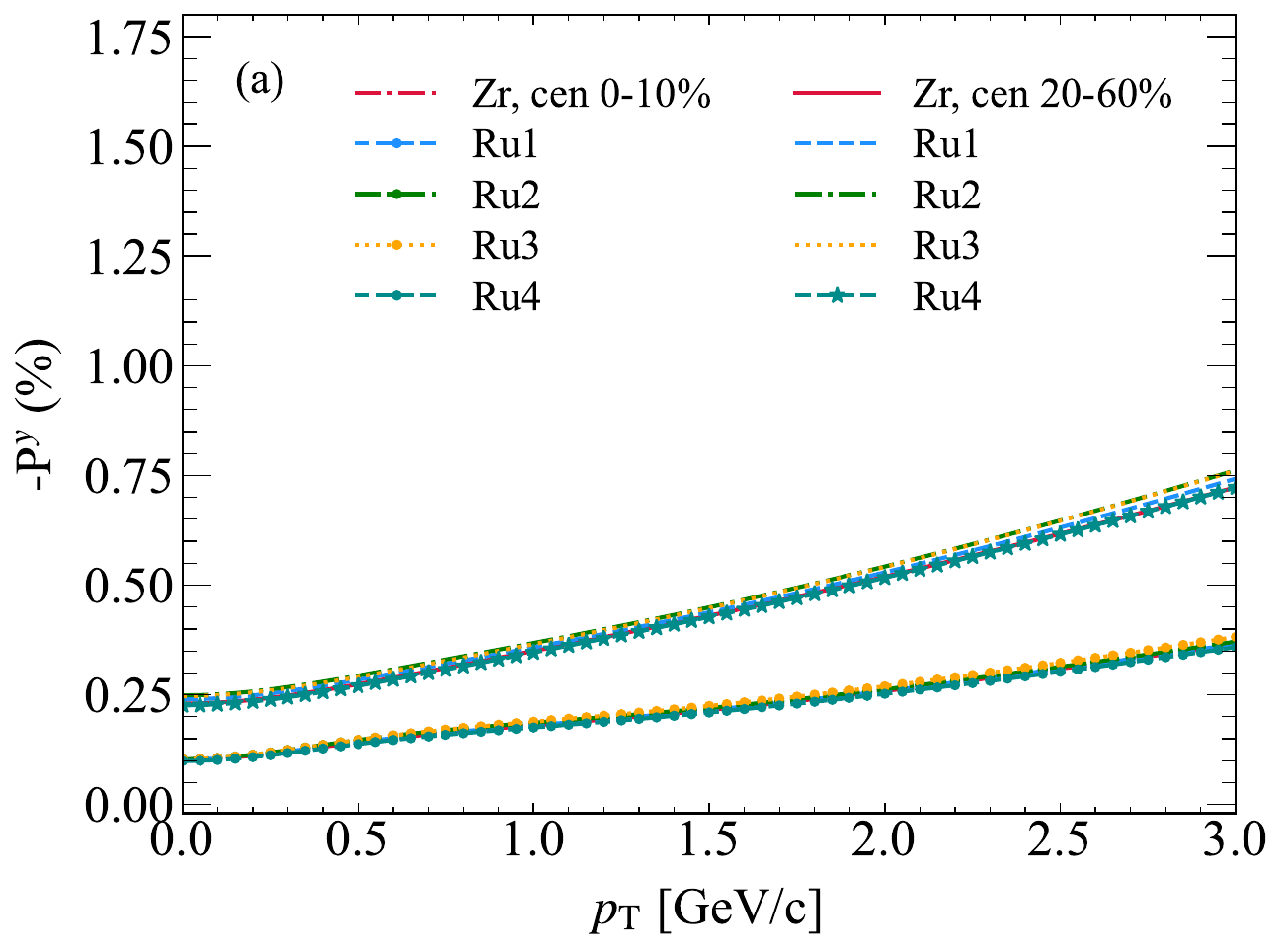} \\
\includegraphics[width=0.85\linewidth]{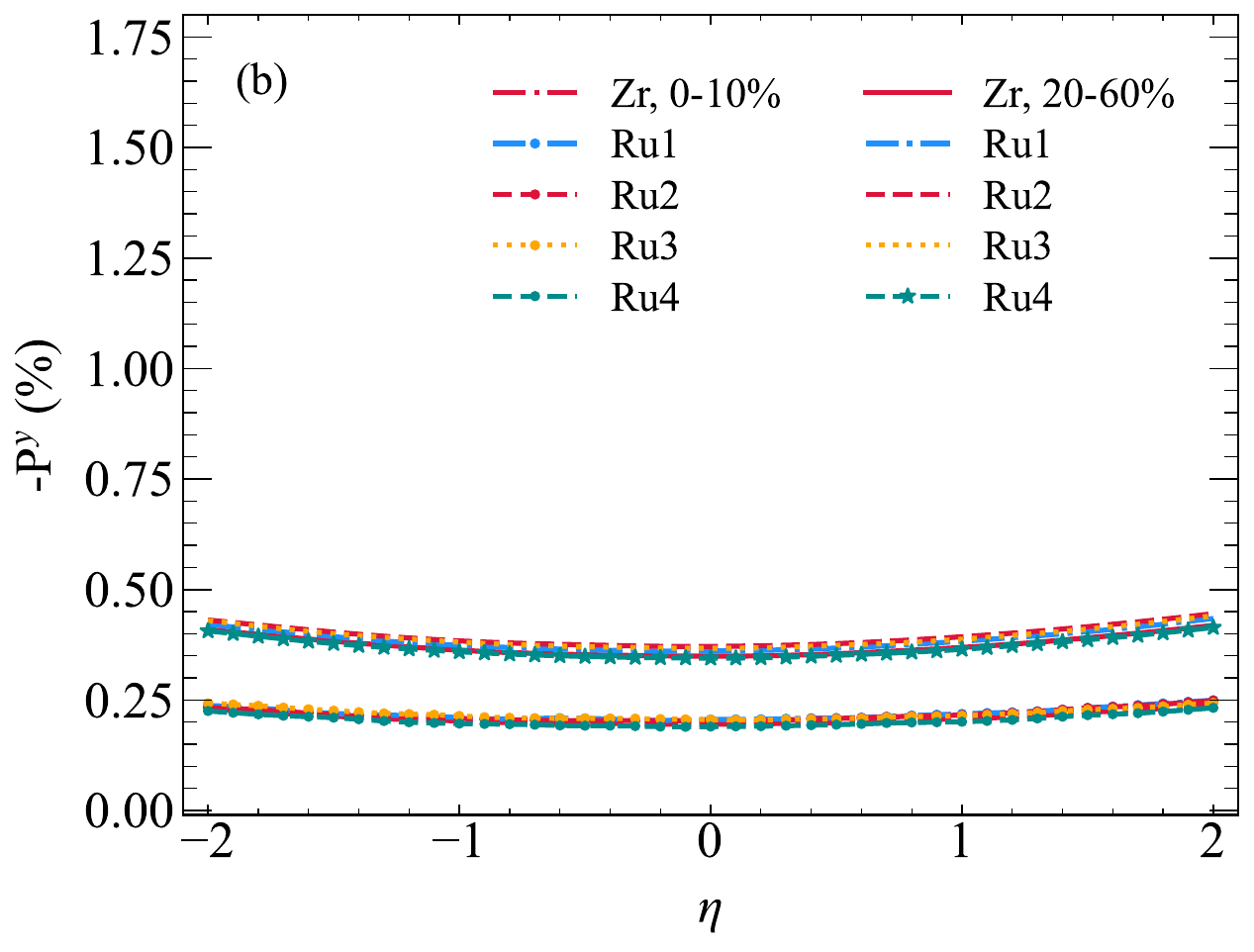}
\end{center}
\caption{(Color online) Dependence of the global $\Lambda$ polarization $-P^{y}$ on
         nuclear structure in the isothermal scenario, at fixed $f_v=0.10$ and
         $k_T=0.33$~GeV. Upper: $-P^{y}$ as a function of $p_T$. Lower: $-P^{y}$ as a
         function of $\eta$. Results for $^{96}$Zr (Case~5, solid lines) and the four
         $^{96}$Ru nuclear structure configurations (Cases~1--4, from
         Table~\ref{tab:zr_structure}) are compared in two centrality bins: 0--10\%
         (lower set of curves) and 20--60\% (upper set).}
\label{f:ns_scan}
\end{figure}

Figure~\ref{f:ns_scan} presents the global polarization obtained with the five nuclear
structure configurations at two representative centralities: 0--10\% and 20--60\%.

The most prominent feature is the near-complete degeneracy of all five configurations
in both the $p_T$ and $\eta$ dependences. In 0--10\% centrality, $-P^{y}$ is small and
the spread among different configurations remains below $0.02\%$ across the entire
plotted $p_T$ range and throughout $|\eta|<1$. In 20--60\% centrality, the polarization
is substantially larger owing to the greater
orbital angular momentum deposited at larger impact parameters, yet the five curves
remain virtually indistinguishable: the maximum deviation between any two nuclear
structures is $\sim 0.03\%$ in the $p_T$-integrated value and $\sim 0.02\%$ in the
$\eta$ dependence, well below the current experimental precision~\cite{STAR:2025dgs}.

This insensitivity starkly contrasts with the strong $f_v$ and $k_T$ dependences,
revealing a clear hierarchy: global polarization is predominantly governed by the total
orbital angular momentum (set by the centrality) and the longitudinal flow profile
(controlled by $f_v$ and, to a lesser extent, $k_T$), whereas the detailed shape of the
nuclear density distribution --- at least within the range $\beta_2\sim 0.06$--$0.16$
and $\beta_3\sim 0.0$--$0.2$ probed by the Zr/Ru isobaric system --- plays a negligible
role. The centrality dependence alone, moving from 0--10\% to 20--60\%, changes
$-P^{y}$ by a factor of approximately 2--3, overwhelmingly dominating over the $<5\%$
variations induced by nuclear structure differences.

This finding is consistent with the STAR observation that $\Lambda$ polarization in
Ru+Ru and Zr+Zr collisions is compatible within experimental
uncertainties~\cite{STAR:2025dgs}, and it supports the use of the combined isobar data
set within the present model uncertainties. From a broader perspective, it indicates
that while global polarization is a powerful probe of the initial flow velocity field
and fireball geometry, it is not an effective observable for constraining nuclear
deformations at the $\beta_2$, $\beta_3 \sim 0.1$ level. Other observables, such as
anisotropic flow ratios~\cite{STAR:2021mii, Jia:2022ozr}, are better suited for that
purpose.

\subsection{\texorpdfstring{Azimuthal angle dependent polarization:
$P_{y,\mathrm{c0}}$ and $P_{y,\mathrm{c2}}$}{Azimuthal angle dependent
polarization: Py,c0 and Py,c2}}
\label{sec:3-3}

Beyond the azimuthal-angle-averaged global polarization discussed in
Secs.~\ref{sec:3-1} and~\ref{sec:3-2}, the dependence of polarization on the hyperon's
azimuthal angle relative to the reaction plane provides a more differential probe of
the polarization mechanism. Experimentally, this dependence is decomposed into Fourier
coefficients~\cite{Niida:2024ntm,STAR:2025dgs}:
\begin{equation}
P_{\Lambda}(\phi_{\rm H}-\Psi_{\rm RP}) \approx P_{y,\mathrm{c0}} +
2\,P_{y,\mathrm{c2}}\cos[2(\phi_{\rm H}-\Psi_{\rm RP})],
\label{eq:Pyc0c2_def}
\end{equation}
where $P_{y,\mathrm{c0}}$ is the constant (azimuth-averaged) term and
$P_{y,\mathrm{c2}}$ quantifies the quadrupole modulation. STAR extracts these
coefficients from the weak decay $\Lambda\to p+\pi^{-}$
($\alpha_{\Lambda}=0.732$~\cite{ParticleDataGroup:2020ssz}) using two complementary
methods (method-1 and method-2) that involve angular correlations between the daughter
proton and the reaction plane, together with the measured $\Lambda$ elliptic flow $v_2$
and detector acceptance corrections~\cite{Niida:2024ntm,STAR:2025dgs}. On the
theoretical side, we compute $P_{y}(\phi_{\rm H}-\Psi_{\rm RP})$ directly from the
local polarization vector and extract $P_{y,\mathrm{c0}}$ and $P_{y,\mathrm{c2}}$ via
Fourier decomposition. To ensure a consistent comparison with the experimental
extraction, we solve the same coupled equations for $P_{y,\mathrm{c0}}$ and
$P_{y,\mathrm{c2}}$ using the model $\Lambda$ elliptic flow $v_2$ as input.

\begin{figure}[tbp!]
\begin{center}
\includegraphics[width=0.85\linewidth]{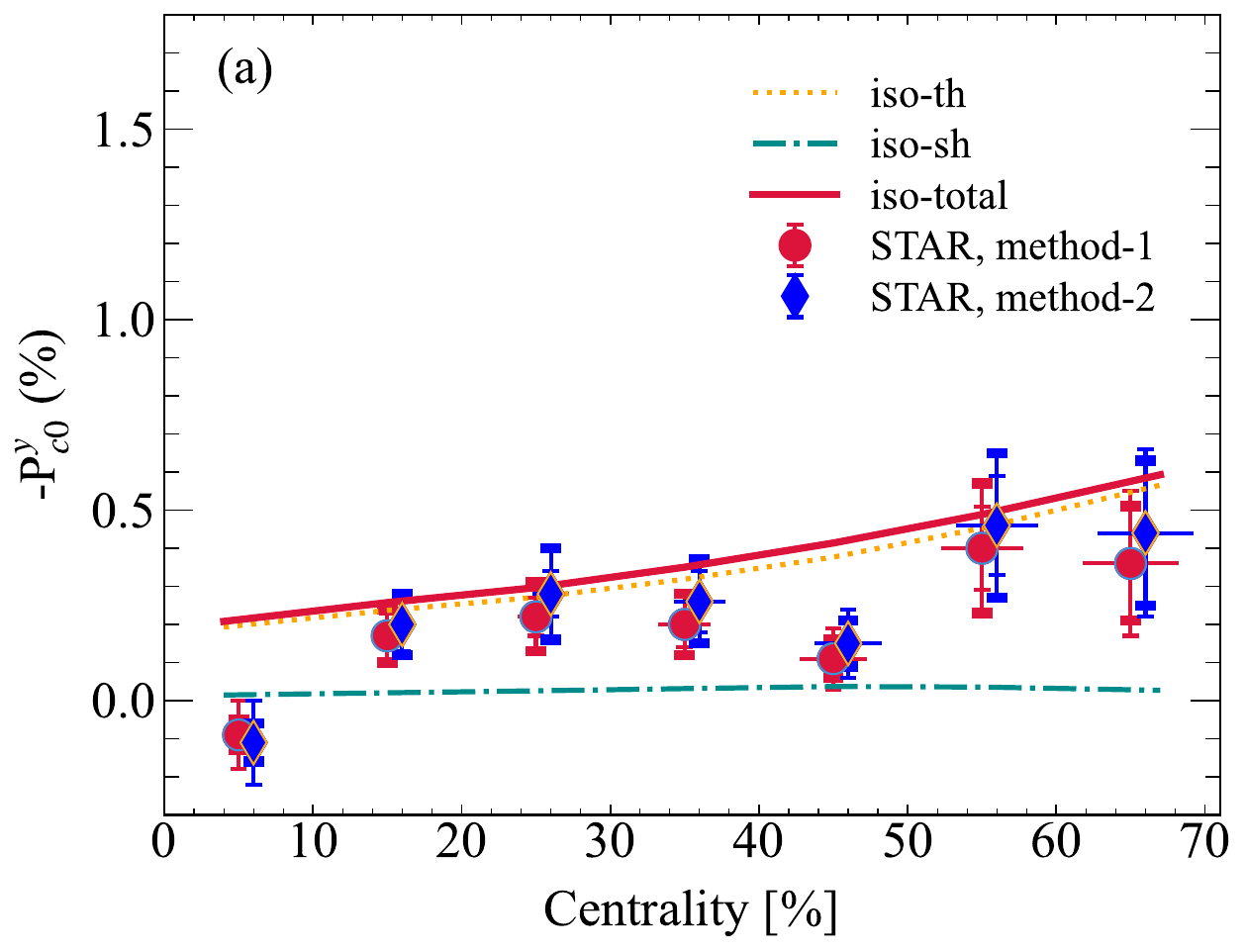} \\
\includegraphics[width=0.85\linewidth]{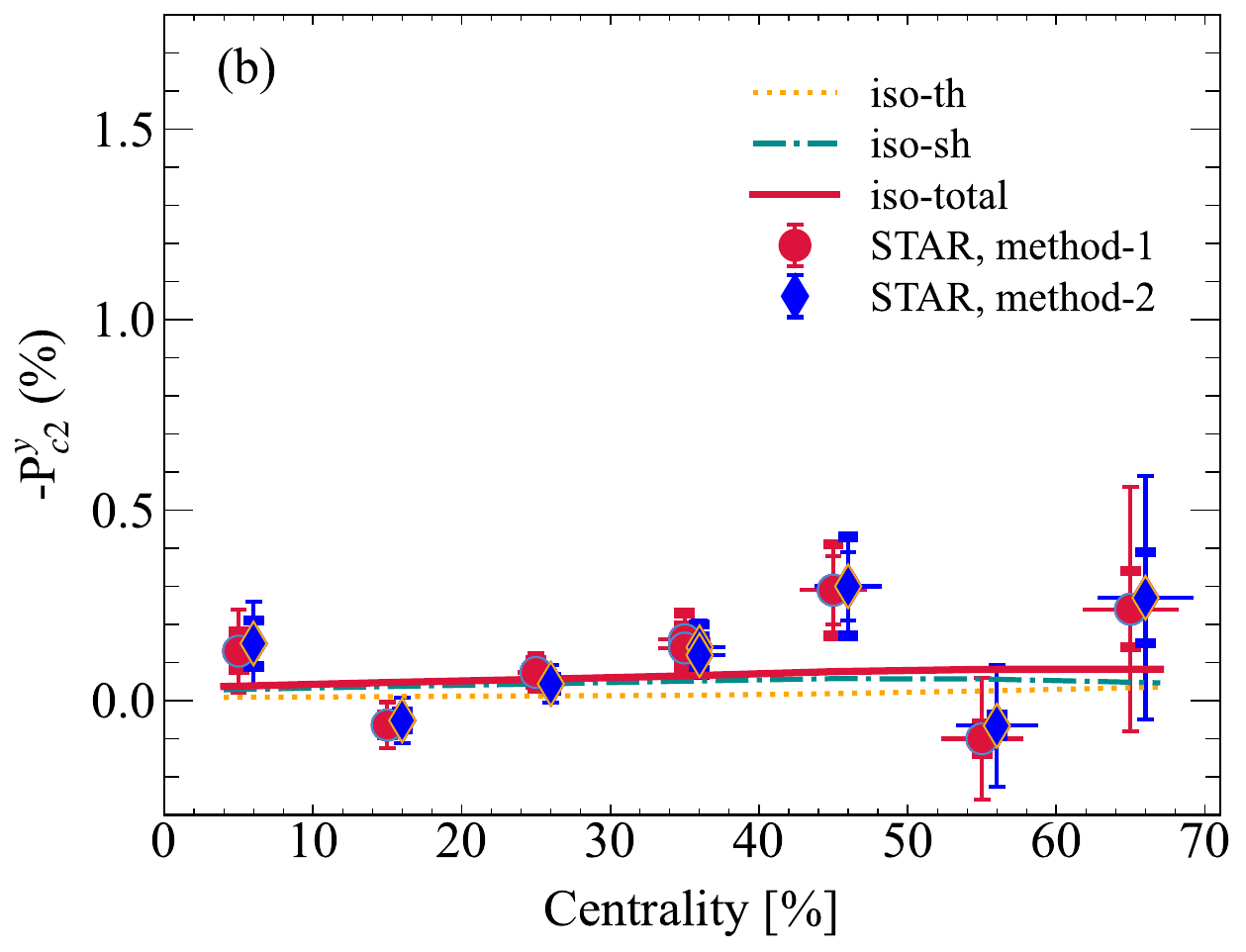}
\end{center}
\caption{(Color online) (a) $-P_{y,\mathrm{c0}}$ and (b) $-P_{y,\mathrm{c2}}$ of
         $\Lambda$ hyperons as functions of centrality in Zr+Zr collisions at
         $\snn=200$~GeV ($p_T\in[0.5,3.0]$~GeV, $|\eta|<1$). The dotted, dashed, and
         solid lines denote `iso-th', `iso-sh', and total isothermal polarization. STAR
         data (method-1: solid circles, method-2: open squares) are for combined Zr+Zr
         and Ru+Ru~\cite{STAR:2025dgs}.}
\label{f:pyc0c2}
\end{figure}

Figure~\ref{f:pyc0c2} presents the centrality dependence of $-P_{y,\mathrm{c0}}$ and
$-P_{y,\mathrm{c2}}$, providing a quantitative theoretical description of these
azimuthal-angle-dependent polarization coefficients in isobaric collisions. The two
coefficients exhibit different sensitivities to the polarization sources. Since the
STAR points combine Zr+Zr and Ru+Ru events, this comparison should be interpreted
together with the nuclear-structure scan in Sec.~\ref{subsubsec-3-3}, which shows that
the predicted polarization is nearly insensitive to the tested isobar parametrizations.

The constant term $-P_{y,\mathrm{c0}}$, shown in panel (a) of Fig.~\ref{f:pyc0c2}, is
closely related to the global polarization $-P^{y}$ and is dominated by the thermal
vorticity contribution across all centralities, while the shear term remains negligible. The total polarization rises from $\sim 0.20\%$ in central collisions to
$\sim 0.55\%$--$0.60\%$ around 60--70\% centrality, in good agreement with both STAR
extraction methods. This confirms that the average polarization magnitude and its
centrality trend are well captured by the model.

A qualitatively different picture emerges for the modulation amplitude
$-P_{y,\mathrm{c2}}$, shown in panel (b) of Fig.~\ref{f:pyc0c2}. The thermal vorticity
contribution is small and nearly independent of centrality,
whereas the shear term grows steadily from $\sim 0.01\%$ to $\sim 0.10\%$ toward
peripheral collisions, becoming the dominant source. Consequently, the total
$-P_{y,\mathrm{c2}}$ is essentially shear-driven over most of the centrality range. The
STAR data, though statistically limited, are broadly consistent with this prediction.

The contrasting behaviors of $P_{y,\mathrm{c0}}$ and $P_{y,\mathrm{c2}}$ reflect their
different physical origins: the former probes the global vorticity of the medium, while
the latter isolates the anisotropic shear
tensor~\cite{Becattini:2021suc,Becattini:2021iol}. The fact that the model
simultaneously reproduces both---with the same parameter set fixed in
Secs.~\ref{sec:trento3d_detail} and~\ref{sec:3-2}---supports the internal consistency
of the isothermal polarization framework.

\begin{figure}[tbp!]
\begin{center}
\includegraphics[width=0.85\linewidth]{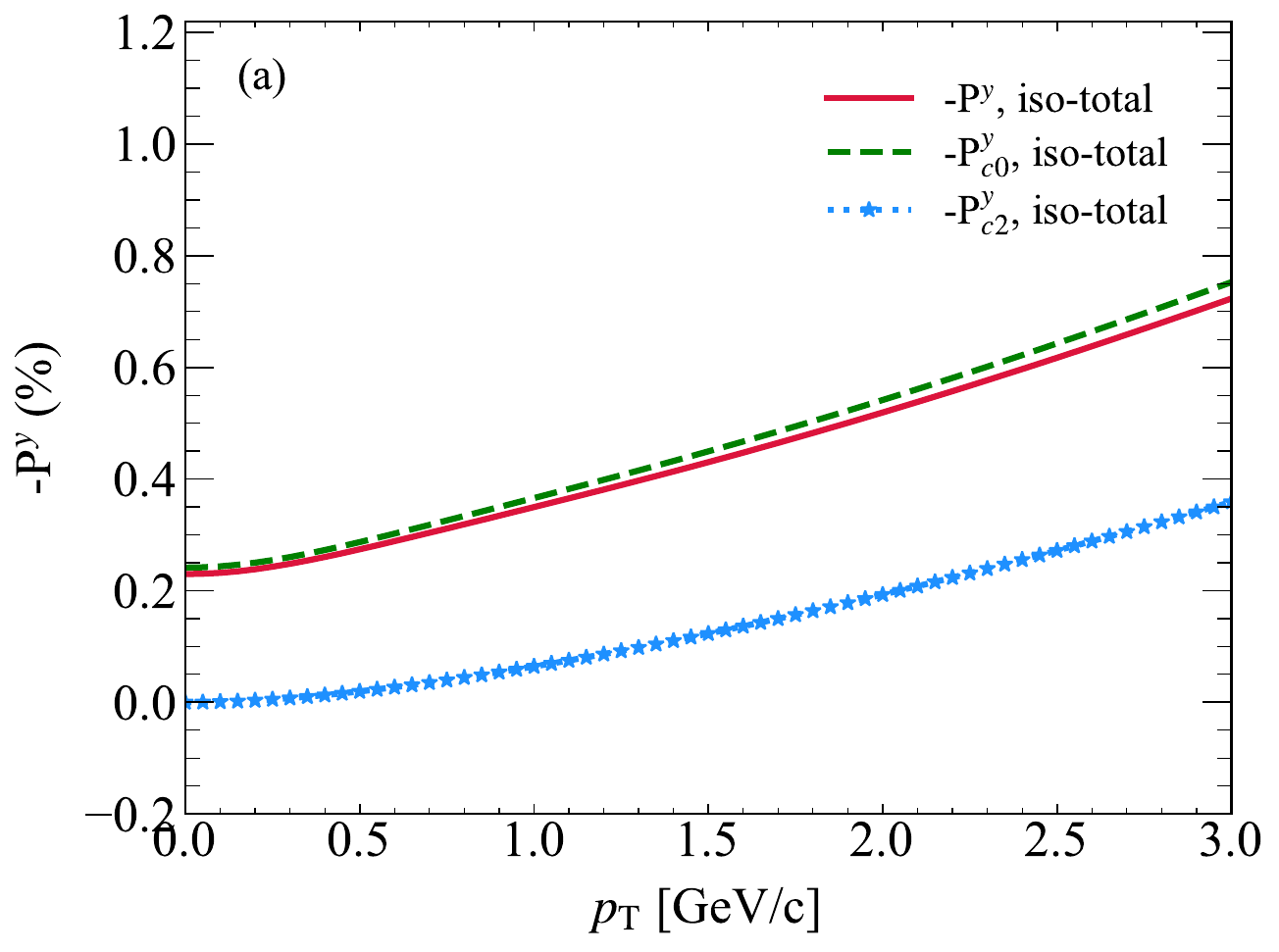}
\end{center}
\caption{(Color online) Theoretical prediction for $-P^{y}$, $-P_{y,\mathrm{c0}}$, and
         $-P_{y,\mathrm{c2}}$ of $\Lambda$ hyperons as functions of $p_T$ in 20--60\%
         Zr+Zr collisions at $\snn=200$~GeV ($|\eta|<1$). The solid, dashed, and dotted
         curves represent the total isothermal polarization, its constant Fourier
         component, and its quadrupole modulation amplitude, respectively.}
\label{f:pyc0c2_pt}
\end{figure}

To further explore the constraining power of the azimuthal decomposition, we present in
Fig.~\ref{f:pyc0c2_pt} our theoretical prediction for the $p_T$ dependence of $-P^{y}$,
$-P_{y,\mathrm{c0}}$, and $-P_{y,\mathrm{c2}}$ in 20--60\% Zr+Zr collisions. The total
polarization $-P^{y}$ (solid) exhibits the characteristic rising $p_T$ dependence
discussed in Sec.~\ref{sec:3-1}, with a magnitude of $\sim 0.75\%$ at high $p_T$. The
constant term $-P_{y,\mathrm{c0}}$ (dashed) closely tracks $-P^{y}$ over the entire
$p_T$ range, albeit at a systematically higher level because it isolates the
azimuthally averaged component of the polarization. The modulation amplitude
$-P_{y,\mathrm{c2}}$ (dotted) is substantially smaller and displays a qualitatively
different $p_T$ dependence: it rises steadily from low to intermediate $p_T$ and
continues to increase at high $p_T$, reflecting the growing importance of the
shear-induced polarization at larger transverse momenta.

The clear separation between $-P_{y,\mathrm{c0}}$ and $-P_{y,\mathrm{c2}}$ in both
magnitude and $p_T$ shape suggests that their measurement as functions of $p_T$ would
provide complementary sensitivity to the thermal-vorticity and shear contributions,
respectively. Such measurements, while statistically challenging, are feasible with the
high-statistics Zr+Zr data samples collected at RHIC and would offer a more
differential constraint on the polarization mechanism than the inclusive $-P^{y}(p_T)$
alone. We propose $-P_{y,\mathrm{c0}}(p_T)$ and $-P_{y,\mathrm{c2}}(p_T)$ as valuable
supplementary observables for probing the vortical and shear structure of the QGP in
heavy-ion collisions.

\subsection{\texorpdfstring{Longitudinal polarization $P_z$ and its azimuthal
modulation}{Longitudinal polarization Pz and its azimuthal modulation}}
\label{sec:3-4}

We now turn to the polarization component along the beam direction, $P_z$, which is
sensitive to the longitudinal gradient of the flow velocity field and provides
complementary information to the out-of-plane component $-P^{y}$ discussed in previous
sections.

\subsubsection{\texorpdfstring{Baseline $P_z$ and comparison with STAR
data}{Baseline Pz and comparison with STAR data}}

\begin{figure}[tbp!]
\begin{center}
\includegraphics[width=0.85\linewidth]{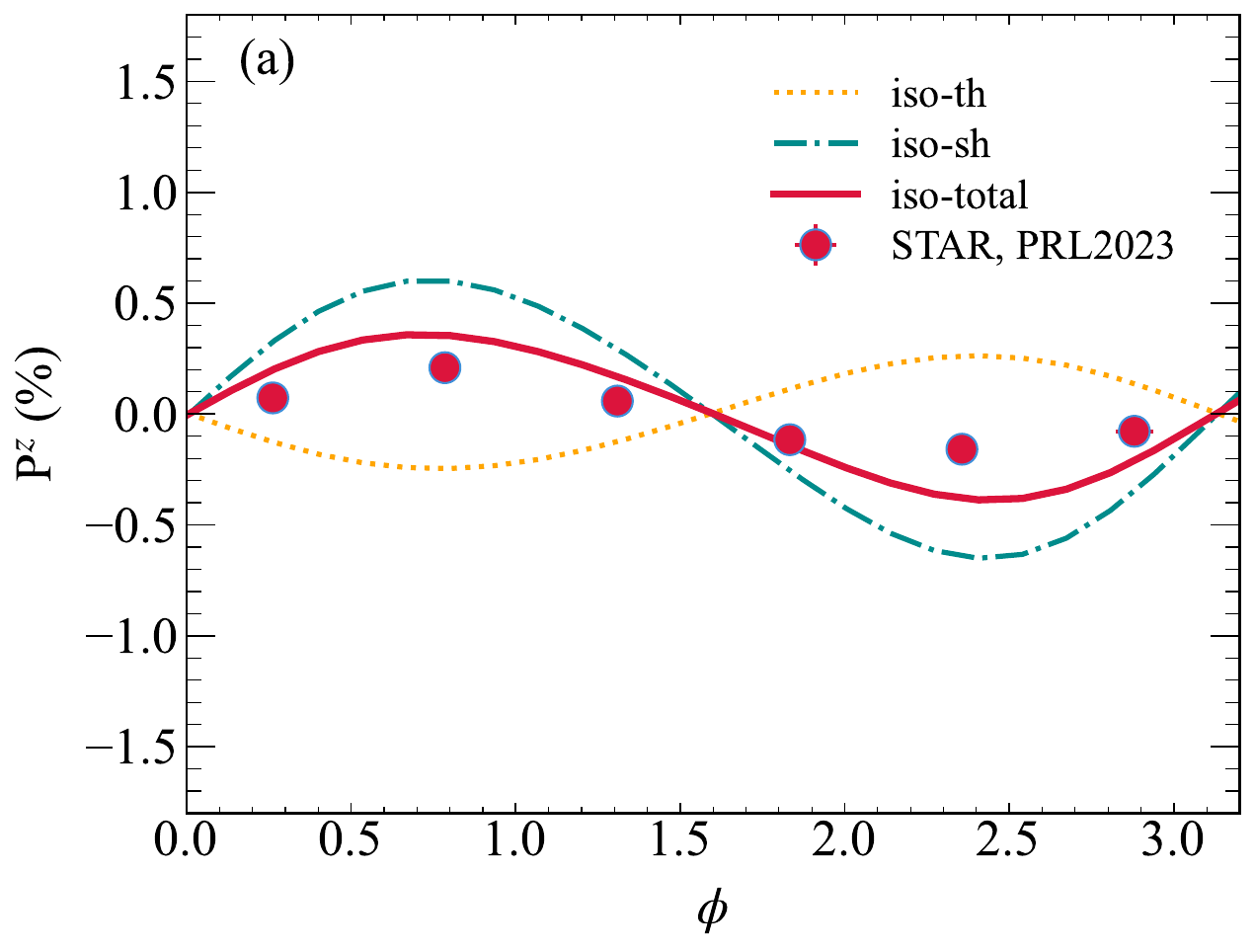} \\
\includegraphics[width=0.85\linewidth]{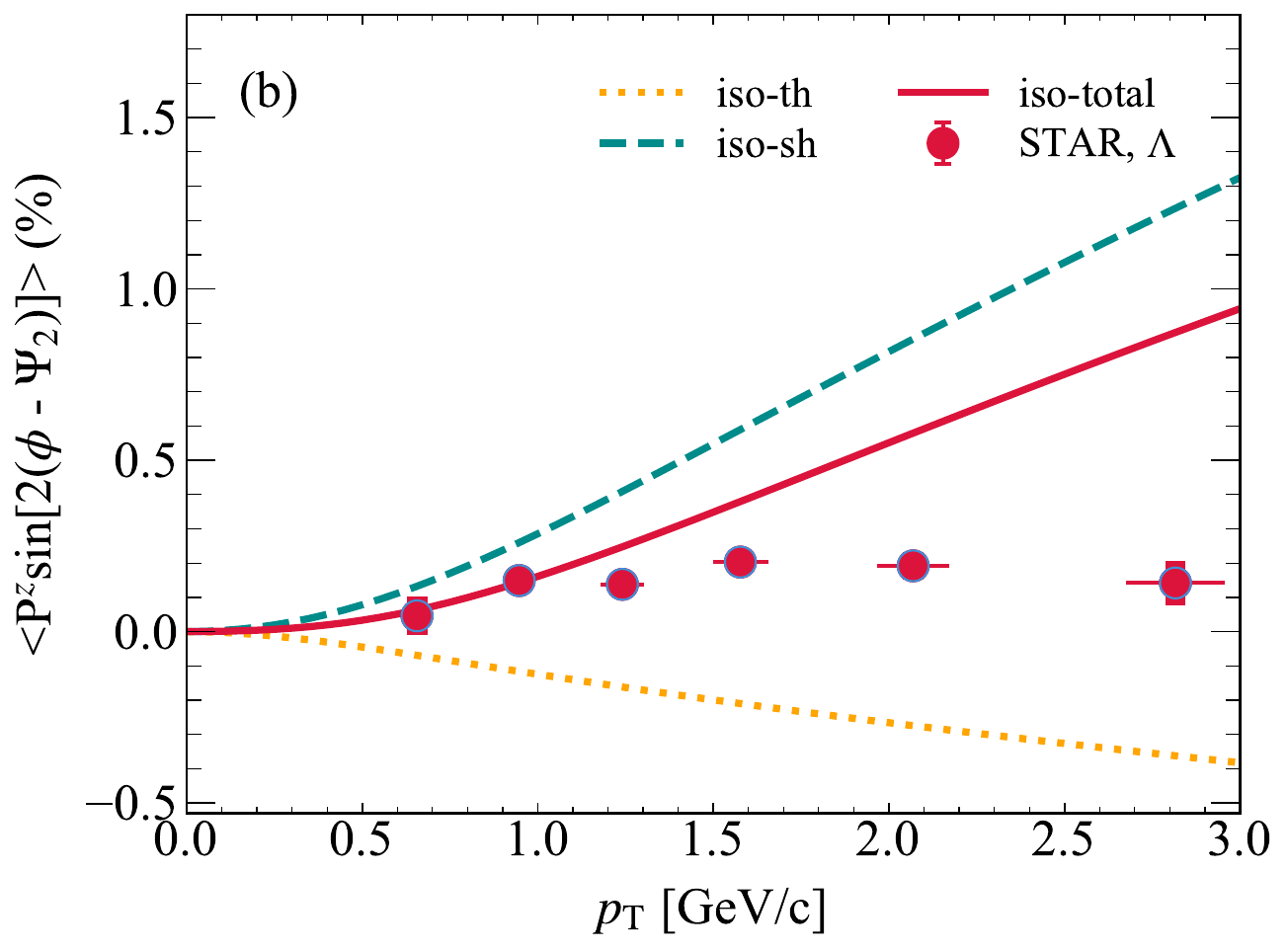}
\end{center}
\caption{(Color online) Longitudinal polarization $P_z$ of $\Lambda$ hyperons in
         20--60\% Zr+Zr collisions at $\snn=200$~GeV ($p_T\in[0.5,3.0]$~GeV,
         $|\eta|<1$). (a) $P_z$ as a function of $\phi-\Psi_2$. The dotted, dashed, and
         solid lines represent `iso-th', `iso-sh', and total isothermal polarization.
         STAR data are from Ref.~\cite{STAR:2023eck}. (b) $\langle
         P_z\sin[2(\phi-\Psi_2)]\rangle$ as a function of $p_T$.}
\label{f:pz_baseline}
\end{figure}

Figure~\ref{f:pz_baseline}(a) presents the azimuthal dependence of $P_z$ in 20--60\%
Zr+Zr collisions. The isothermal vorticity contribution (`iso-th') produces a negative
sinusoidal modulation, while the shear contribution (`iso-sh') yields a stronger
positive sinusoidal pattern. Their competition results in a net modulation whose sign
and amplitude are determined by the relative strength of the two components. The total
polarization agrees reasonably well with the STAR measurements~\cite{STAR:2023eck},
showing that the isothermal framework captures the observed azimuthal phase in this
isobaric system.

Figure~\ref{f:pz_baseline} Panel (b) shows the $p_T$ dependence of the second Fourier sine coefficient $\langle
P_z\sin[2(\phi-\Psi_2)]\rangle$, which quantifies the leading azimuthal modulation of
the longitudinal polarization. This observable exhibits a strong and nearly monotonic
rise with $p_T$ in our calculation. The shear contribution dominates at all $p_T$ and
grows particularly rapidly at high $p_T$, while the thermal vorticity term remains
moderate throughout. The STAR data, however, show a much weaker $p_T$ dependence,
remaining below $\sim 0.3\%$ up to $p_T=3$~GeV. Consequently, while the model agrees
with the data at low $p_T$ ($\lesssim 1.2$~GeV), it substantially overpredicts the
measurement at intermediate and high $p_T$.

This discrepancy is not unique to our framework. A qualitatively similar overprediction
of the high-$p_T$ $\langle P_z\sin[2(\phi-\Psi_2)]\rangle$ was observed by the
hydrodynamic study of Alzhrani~\emph{et al.}~\cite{Alzhrani:2022dpi,STAR:2023eck},
suggesting that additional physical mechanisms not yet included in the current model
may be required to suppress the longitudinal polarization at high transverse momenta.
Among the potential candidates are the effects of bulk viscosity, electromagnetic
fields, the acceleration-gradient-induced vorticity term, and late-stage hadronic
effects. We leave these investigations to future work.

\subsubsection{\texorpdfstring{Parameter dependence of $P_z$}{Parameter dependence of Pz}}

We next examine how the longitudinal polarization responds to variations of $f_v$,
$k_T$, nuclear structure, and bulk viscosity. For readability, we discuss these scans
in the same order as in the global-polarization analysis: first the longitudinal flow
fraction, then the transverse momentum scale, followed by nuclear structure and
bulk-viscosity effects.

\begin{figure}[tbp!]
\begin{center}
\includegraphics[width=0.85\linewidth]{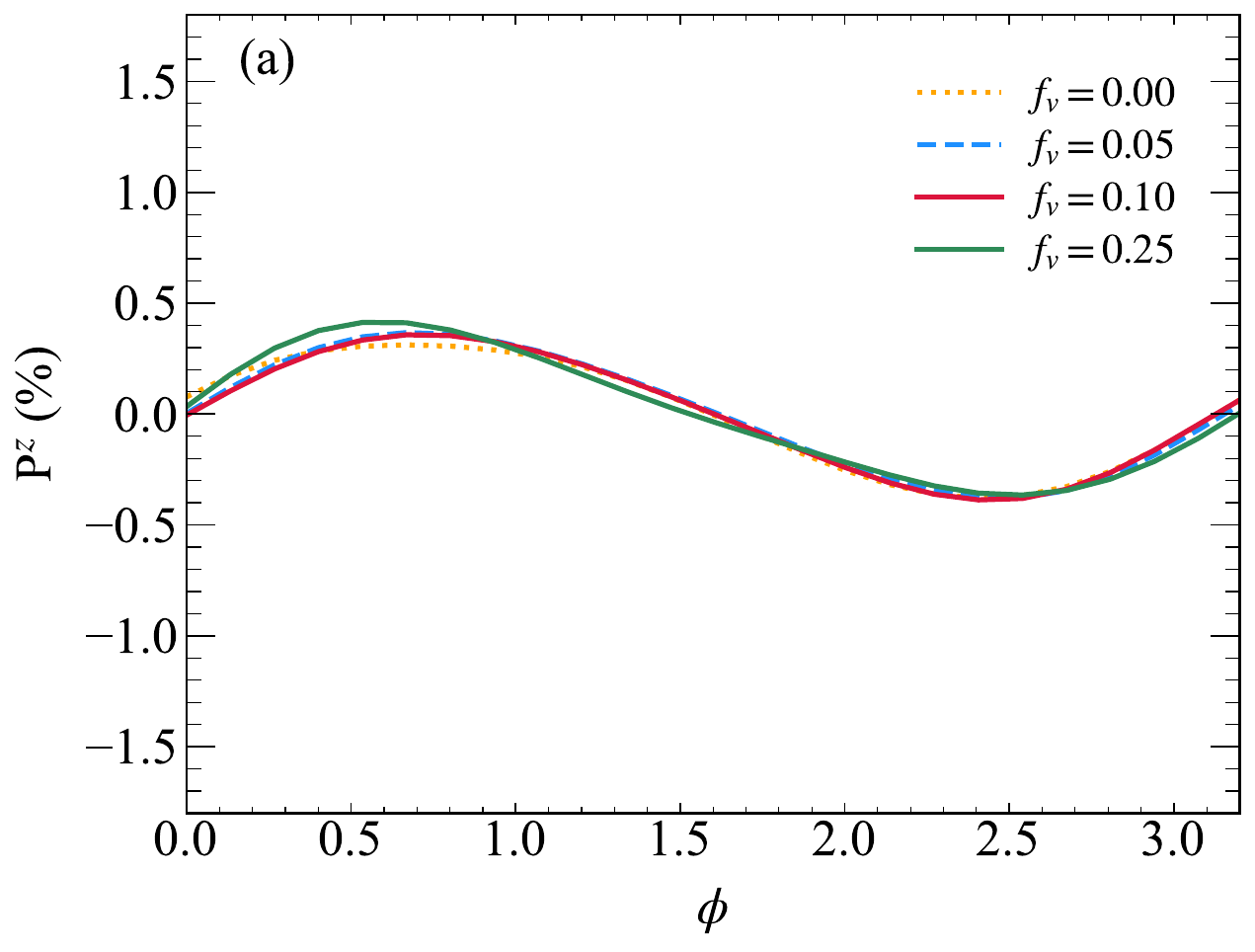} \\
\includegraphics[width=0.85\linewidth]{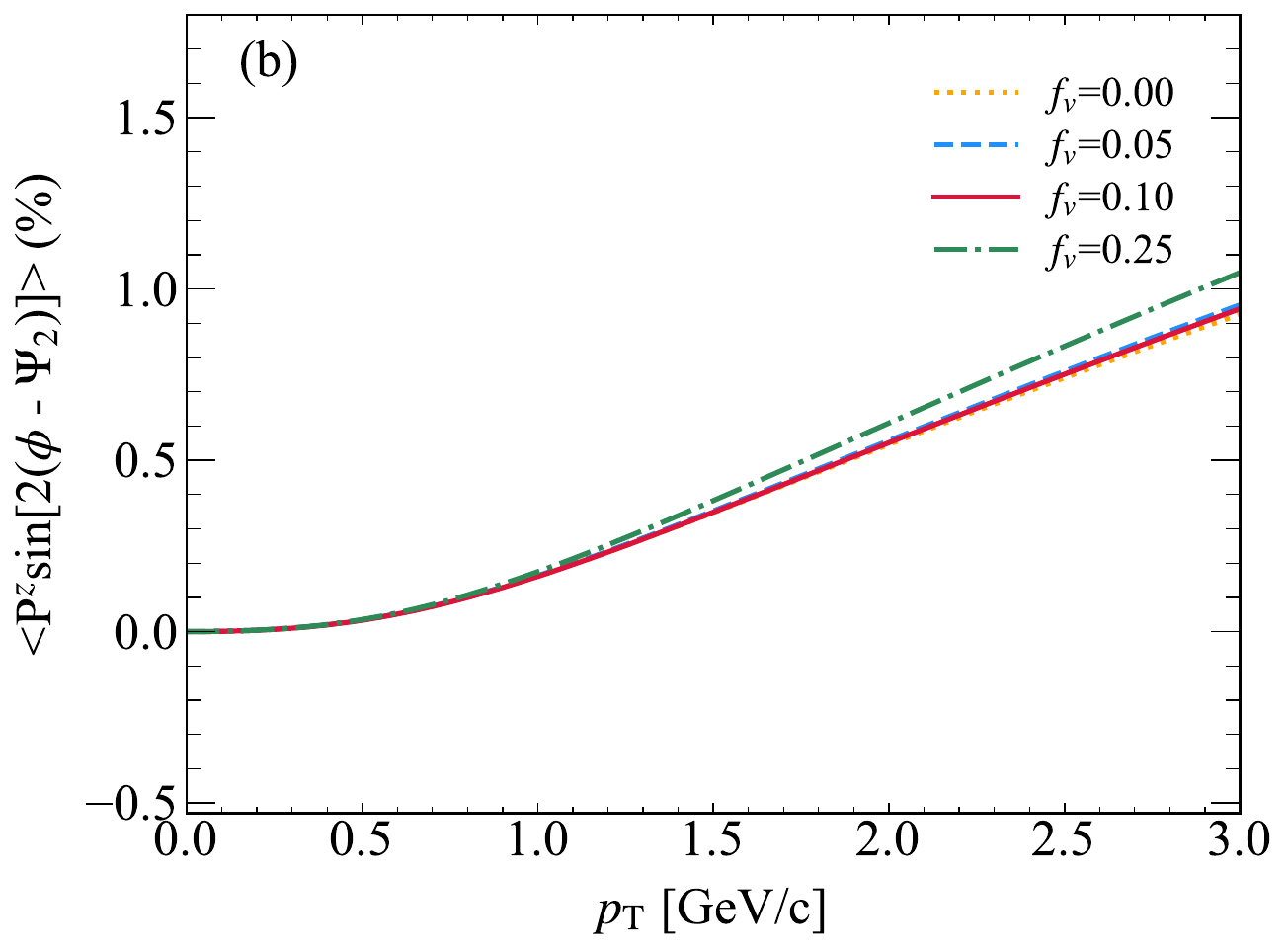}
\end{center}
\caption{(Color online) Dependence of $P_z$ on the longitudinal flow fraction $f_v$ in
         20--60\% Zr+Zr collisions. (a) $P_z(\phi)$. (b) $\langle
         P_z\sin[2(\phi-\Psi_2)]\rangle(p_T)$.}
\label{f:pz_fv}
\end{figure}

Figure~\ref{f:pz_fv} examines the $f_v$ dependence of $P_z$. Panel (a) shows that
$P_z(\phi)$ is remarkably insensitive to $f_v$: all four curves ($f_v=0.00$ to $0.25$)
exhibit a nearly identical negative sinusoidal modulation. Even in the Bjorken limit
$f_v=0$, $P_z(\phi)$ maintains a finite amplitude ($\sim 0.4\%$--$0.5\%$), indicating
that the tilted fireball geometry alone can generate longitudinal polarization---in
sharp contrast to the strong $f_v$ dependence of $-P^{y}$ (Fig.~\ref{f:fv_scan}). Panel
(b), however, reveals a slightly different behavior for the $p_T$-differential
modulation $\langle P_z\sin[2(\phi-\Psi_2)]\rangle$. At $f_v=0$, this observable grows
mildly with $p_T$, while at larger $f_v$ it increases a little at high $p_T$, driven by
the enhanced shear contribution from the stronger longitudinal velocity gradient. The
$\phi$-integrated $P_z$ amplitude is thus nearly $f_v$-blind, yet its $p_T$-decomposed
modulation is weakly $f_v$-sensitive. This contrasting sensitivity provides a
constraint on the longitudinal flow gradient.

\begin{figure}[tbp!]
\begin{center}
\includegraphics[width=0.85\linewidth]{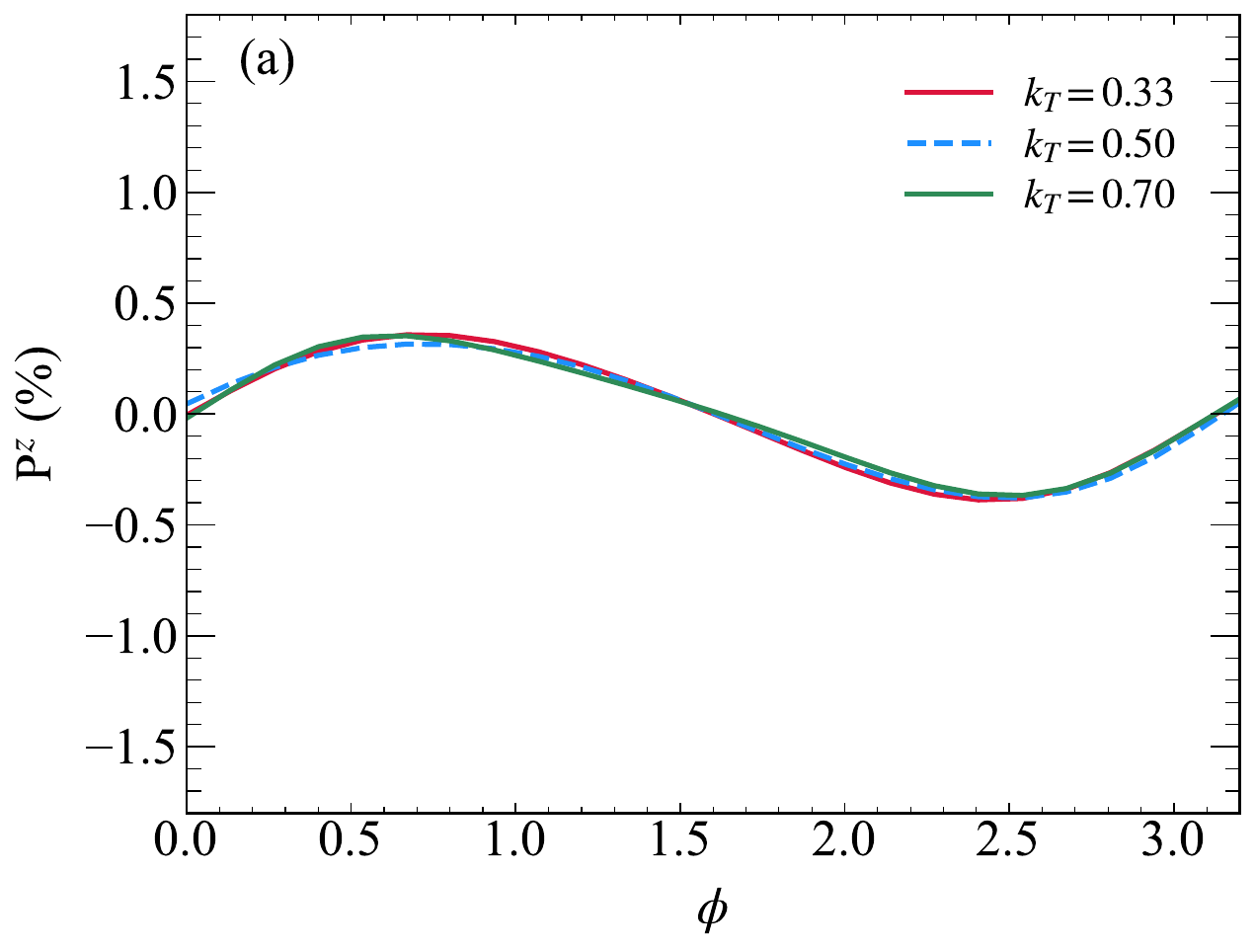} \\
\includegraphics[width=0.85\linewidth]{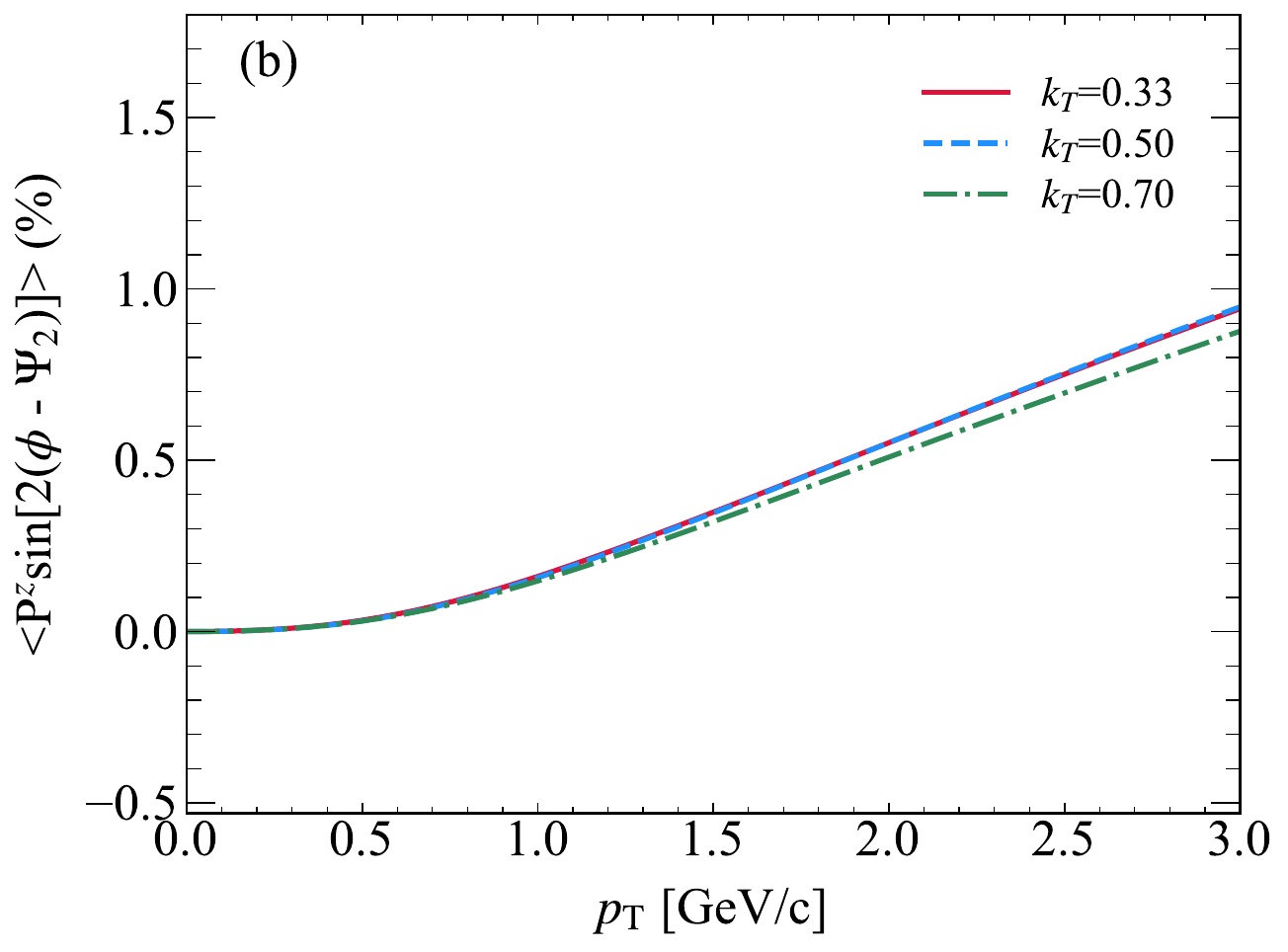}
\end{center}
\caption{(Color online) Dependence of $P_z$ on the transverse momentum scale $k_T$ in
         20--60\% Zr+Zr collisions. (a) $P_z(\phi)$. (b) $\langle
         P_z\sin[2(\phi-\Psi_2)]\rangle(p_T)$.}
\label{f:pz_kt}
\end{figure}

Figure~\ref{f:pz_kt} presents the $k_T$ dependence of $P_z$. Both the azimuthal
modulation $P_z(\phi)$ in panel (a) and its $p_T$-differential amplitude $\langle
P_z\sin[2(\phi-\Psi_2)]\rangle(p_T)$ in panel (b) are strikingly insensitive to $k_T$:
the three curves ($k_T=0.33$, 0.50, and 0.70~GeV) are nearly indistinguishable across
the full $\phi$ and $p_T$ ranges. This null dependence, together with the similarly
weak $f_v$ sensitivity of $P_z(\phi)$ shown in Fig.~\ref{f:pz_fv}(a), indicates that
the longitudinal polarization is driven primarily by the flow velocity field rather
than by the geometric tilt of the fireball, consistent with earlier hydrodynamic
studies~\cite{Jiang:2023vxp,Palermo:2024tza}. While the $p_T$-differential modulation
shows some $f_v$ dependence at high $p_T$ [Fig.~\ref{f:pz_fv}(b)], the persistent
overprediction of $\langle P_z\sin[2(\phi-\Psi_2)]\rangle$, unresolved by tuning either
$f_v$ or $k_T$, further points to additional mechanisms, such as the bulk-viscosity
effects discussed in the next subsection.

\begin{figure}[tbp!]
\begin{center}
\includegraphics[width=0.85\linewidth]{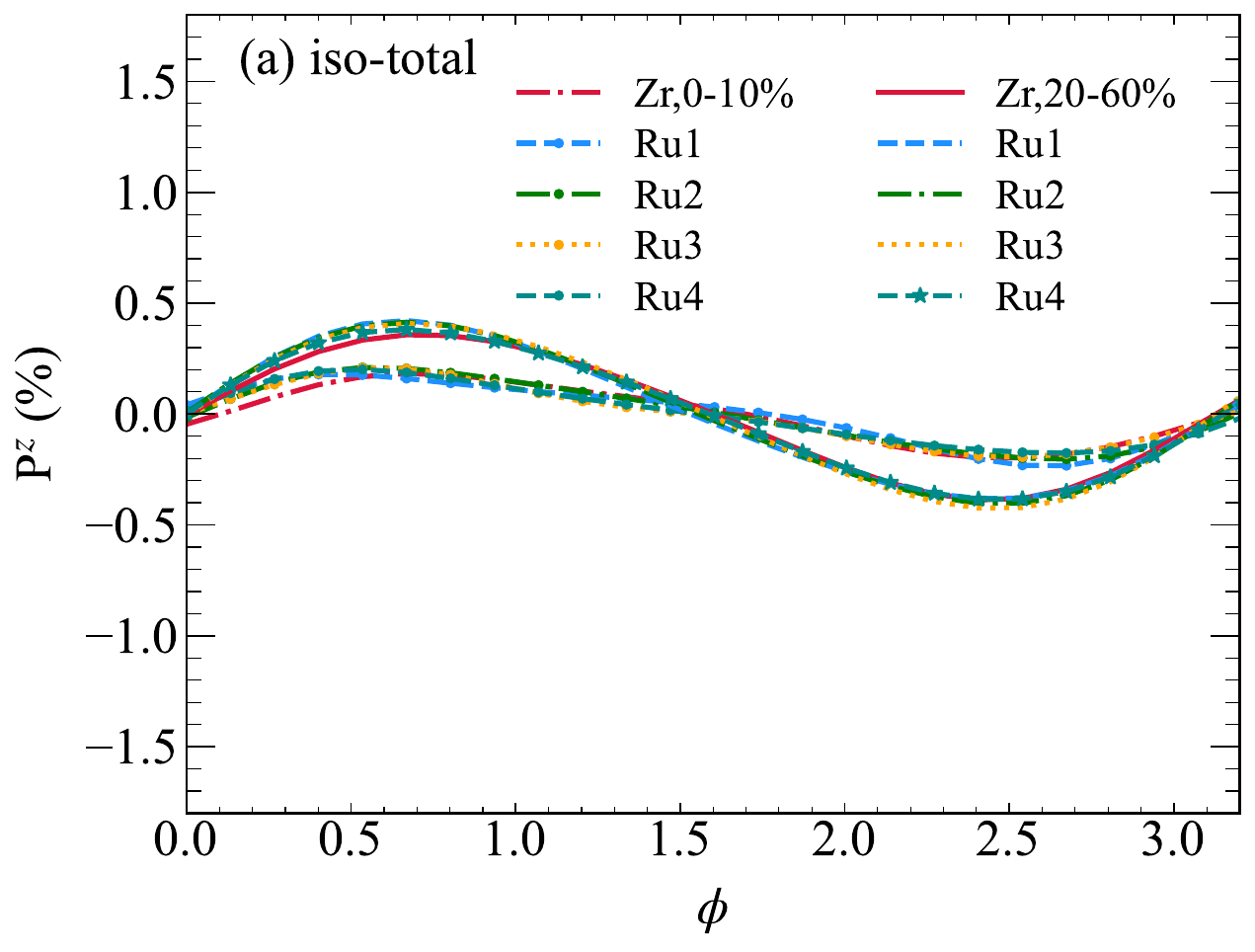} \\
\includegraphics[width=0.85\linewidth]{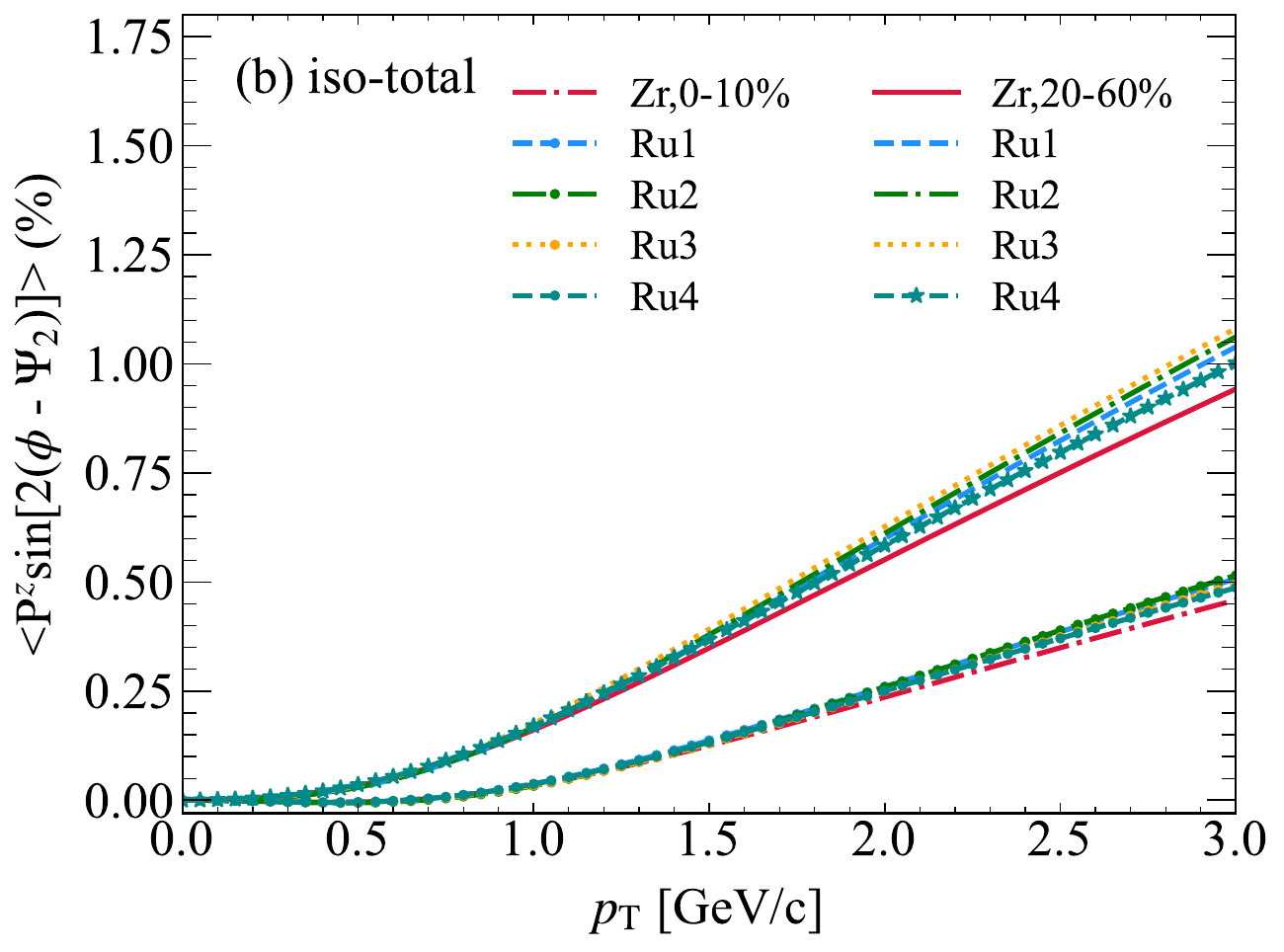}
\end{center}
\caption{(Color online) Dependence of $P_z$ on nuclear structure in 0-10\% and 20--60\%
         Zr+Zr collisions. (a) $P_z(\phi)$. (b) $\langle
         P_z\sin[2(\phi-\Psi_2)]\rangle(p_T)$. Five configurations are compared:
         $^{96}$Zr and four $^{96}$Ru nuclear structure configurations
         (Table~\ref{tab:zr_structure}).}
\label{f:pz_ns}
\end{figure}

Figure~\ref{f:pz_ns} compares $P_z$ across the five nuclear structure configurations in
0-10\% and 20--60\% Zr+Zr collisions. Consistent with the findings for $-P^{y}$
(Sec.~\ref{sec:3-2}), the curves for all configurations are virtually coincident in
both $P_z(\phi)$ and $\langle P_z\sin[2(\phi-\Psi_2)]\rangle(p_T)$. This reaffirms the
conclusion that global and local polarization observables are insensitive to nuclear
deformations at the level probed by the Zr/Ru isobaric system.

\subsubsection{Impact of bulk viscosity}

\begin{figure}[tbp!]
\begin{center}
\includegraphics[width=0.85\linewidth]{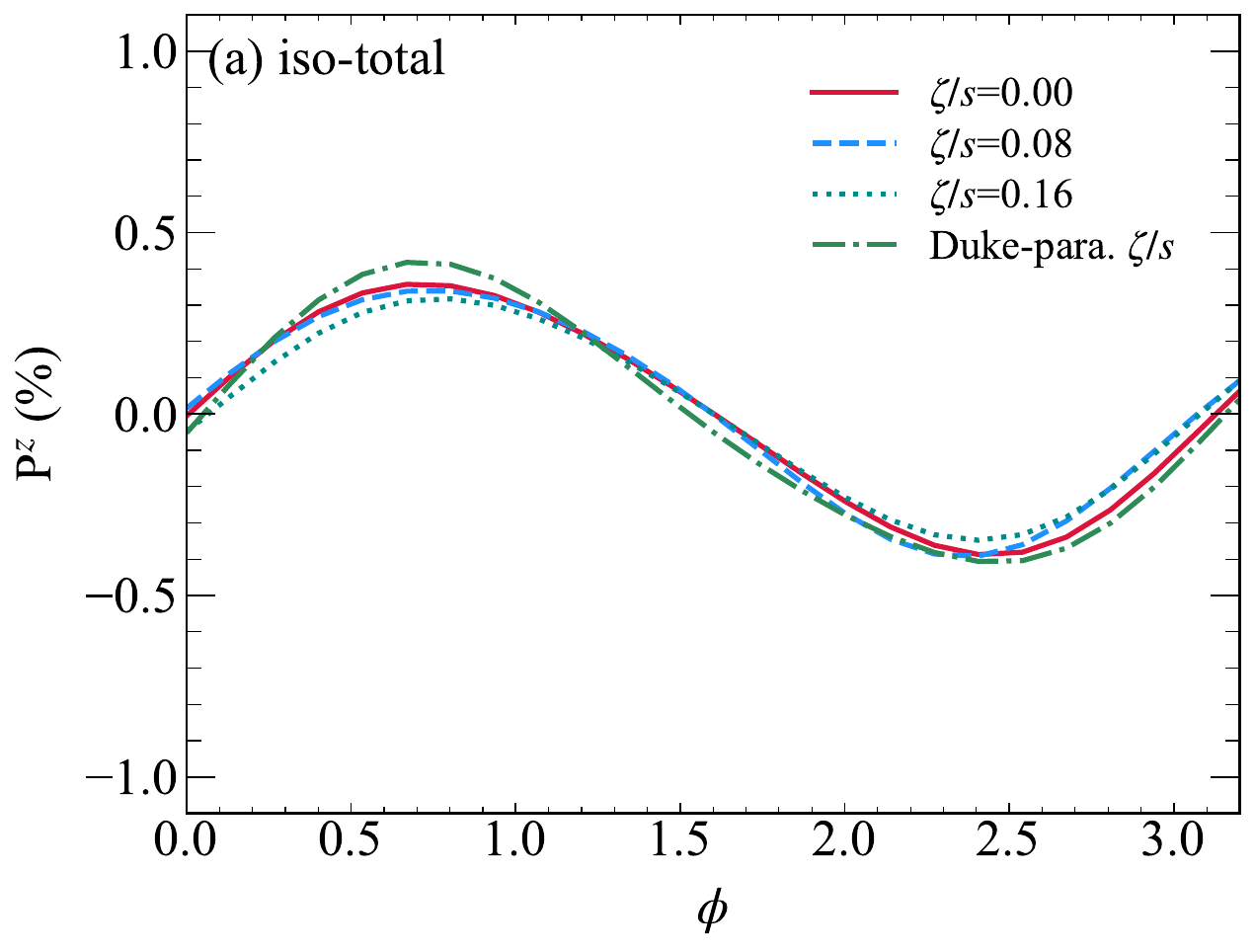} \\
\includegraphics[width=0.85\linewidth]{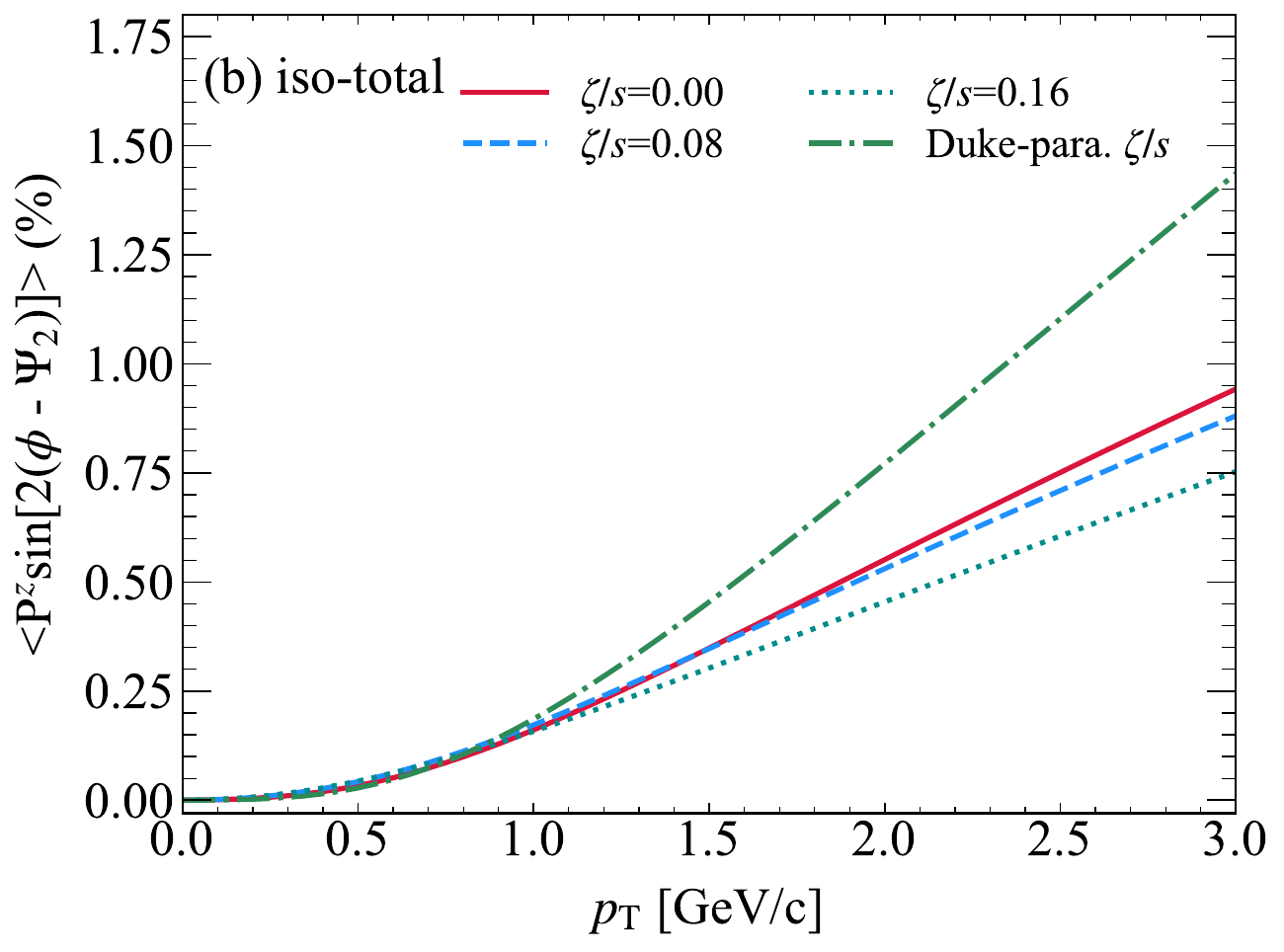}
\end{center}
\caption{(Color online) Dependence of $P_z$ on the specific bulk viscosity $\zeta/s$ in
         20--60\% Zr+Zr collisions. (a) $P_z(\phi)$. (b) $\langle
         P_z\sin[2(\phi-\Psi_2)]\rangle(p_T)$. Three constant values
         $\zeta/s=0.00,0.08,0.16$ and the Duke temperature-dependent parametrization
         [Eq.~(\ref{eq:zetas})] are compared.}
\label{f:pz_bulk}
\end{figure}

Bulk viscosity is expected to damp the expansion rate and thus suppress the shear
tensor that dominates $P_z$, potentially reducing the longitudinal polarization at high
$p_T$. Figure~\ref{f:pz_bulk} examines this effect by comparing results for
$\zeta/s=0.00$, 0.08, 0.16, and the Duke temperature-dependent
parametrization~\cite{Ryu:2017qzn}. Panel (a) shows that increasing $\zeta/s$ mildly
reduces the amplitude of $P_z(\phi)$.

The impact on the $p_T$ dependence shown in panel (b) is more pronounced, but the sign of the
effect depends on the parametrization. The constant bulk viscosities $\zeta/s=0.08$ and
$0.16$ moderately suppress $\langle P_z\sin[2(\phi-\Psi_2)]\rangle$ at high $p_T$
relative to the inviscid case: at $p_T=3$~GeV, the modulation amplitude drops from
$\sim 0.90\%$ at $\zeta/s=0$ to $\sim 0.75\%$ at $\zeta/s=0.16$. In contrast, the
temperature-dependent Duke parametrization \emph{enhances} the modulation, yielding a
value of $\sim 1.40\%$---significantly larger than all constant $\zeta/s$ results. This
mixed behavior indicates that bulk viscosity, at least within the range constrained by
soft hadron observables~\cite{Ryu:2017qzn,Bernhard:2019bmu}, cannot resolve the
high-$p_T$ overprediction of $\langle P_z\sin[2(\phi-\Psi_2)]\rangle$; the Duke form,
in fact, exacerbates the discrepancy. Additional mechanisms beyond bulk viscosity are
therefore required.

\begin{figure*}[tbp!]
\begin{center}
\includegraphics[width=0.48\linewidth]{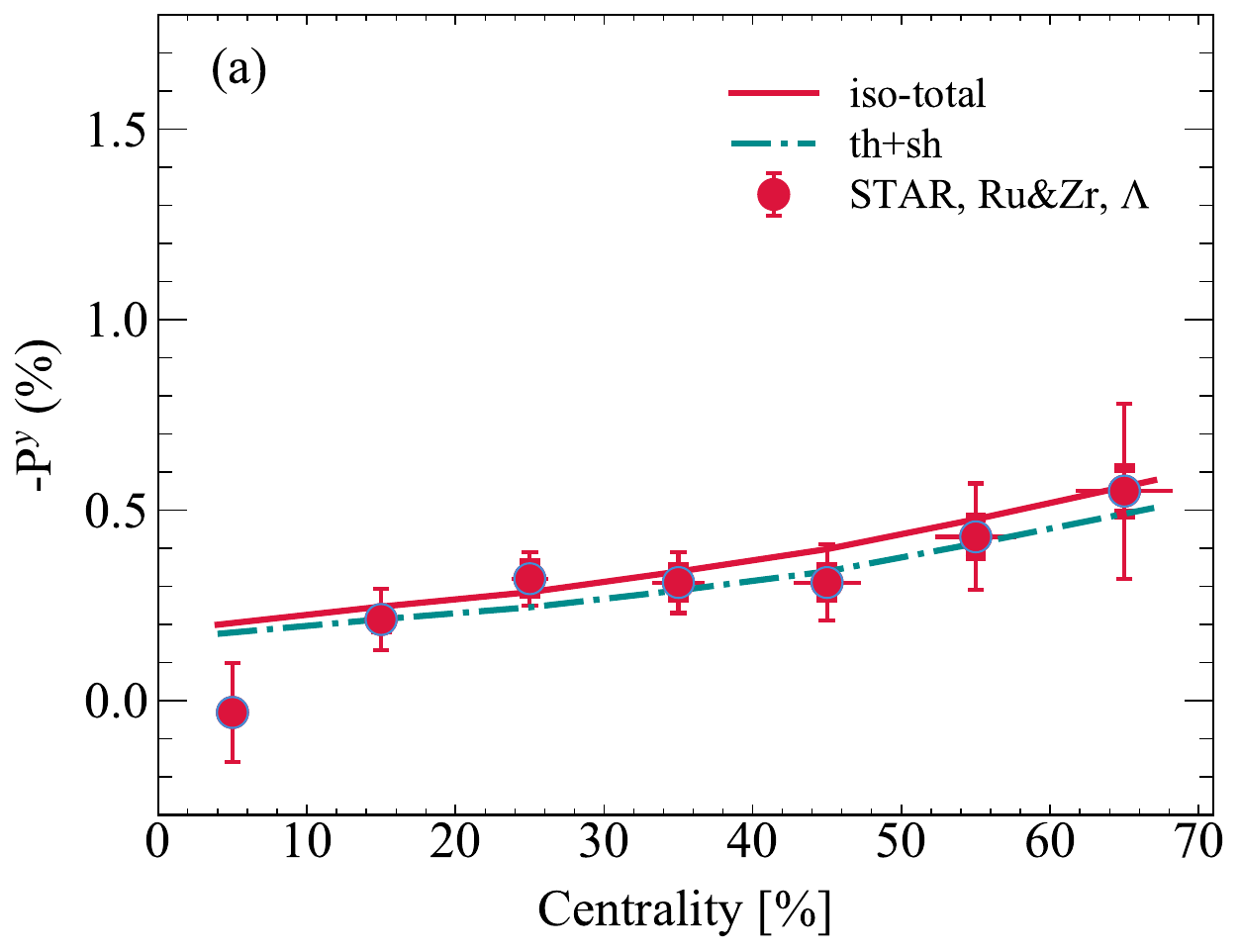}~~
\includegraphics[width=0.48\linewidth]{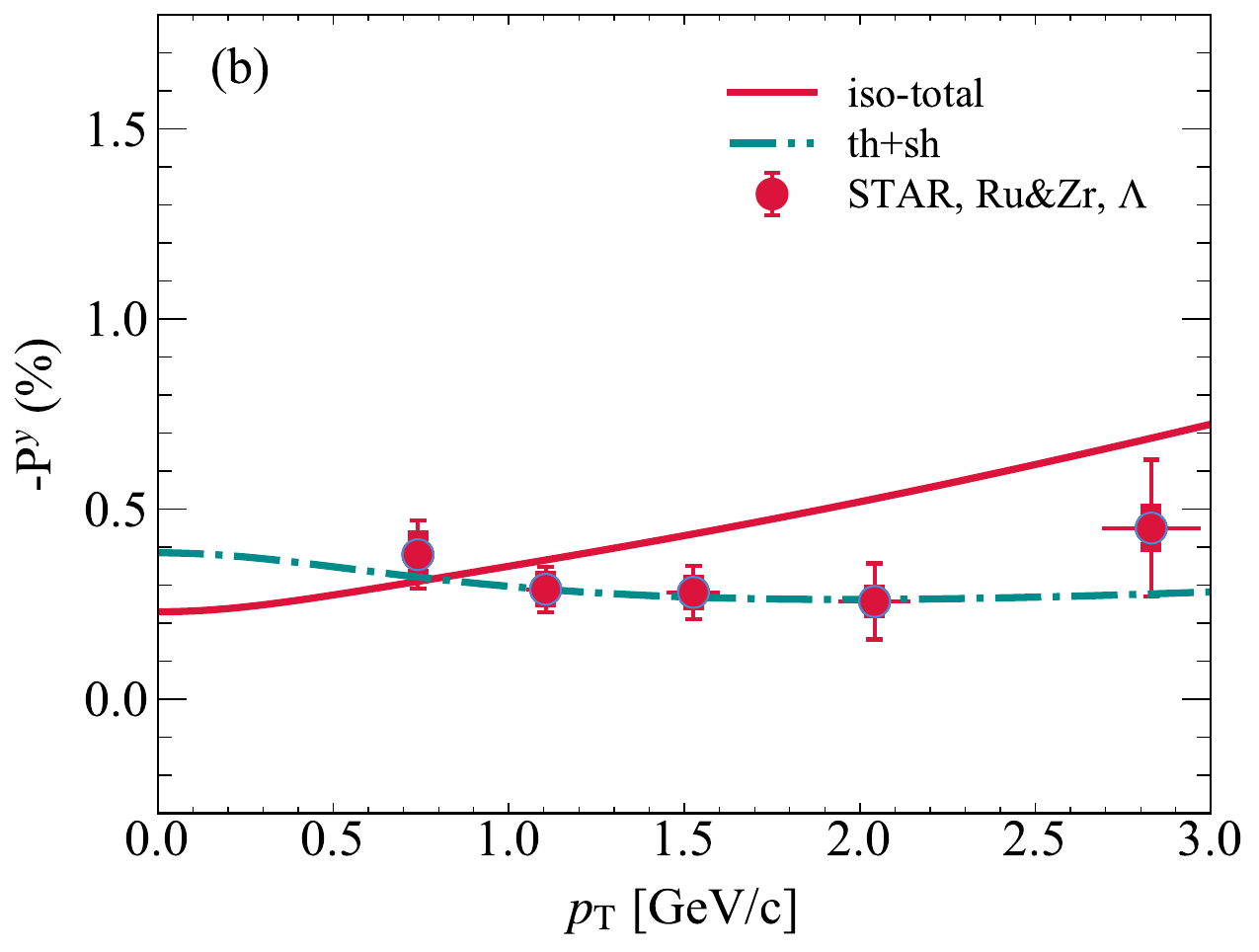} \\
\includegraphics[width=0.48\linewidth]{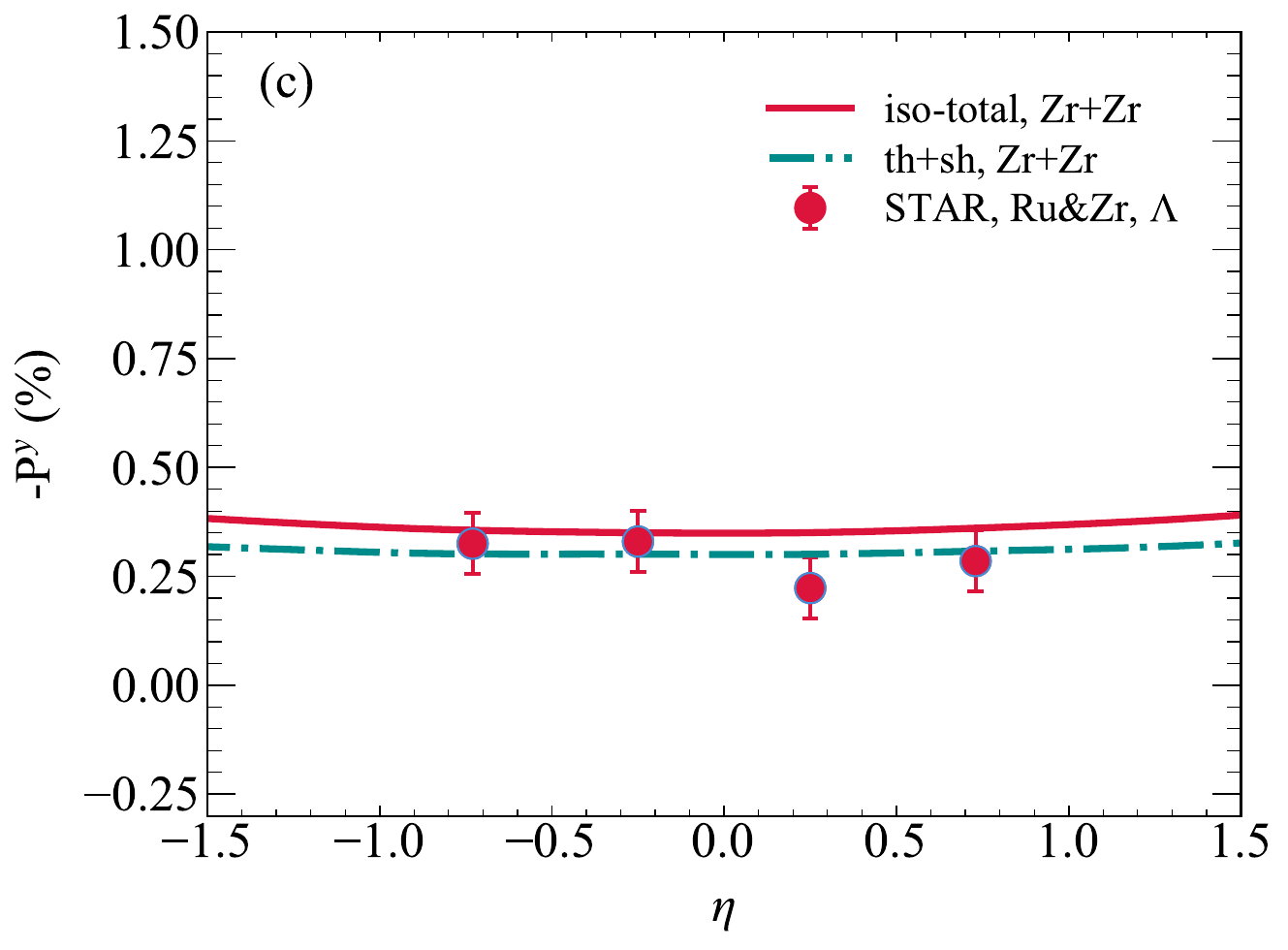}~~
\includegraphics[width=0.48\linewidth]{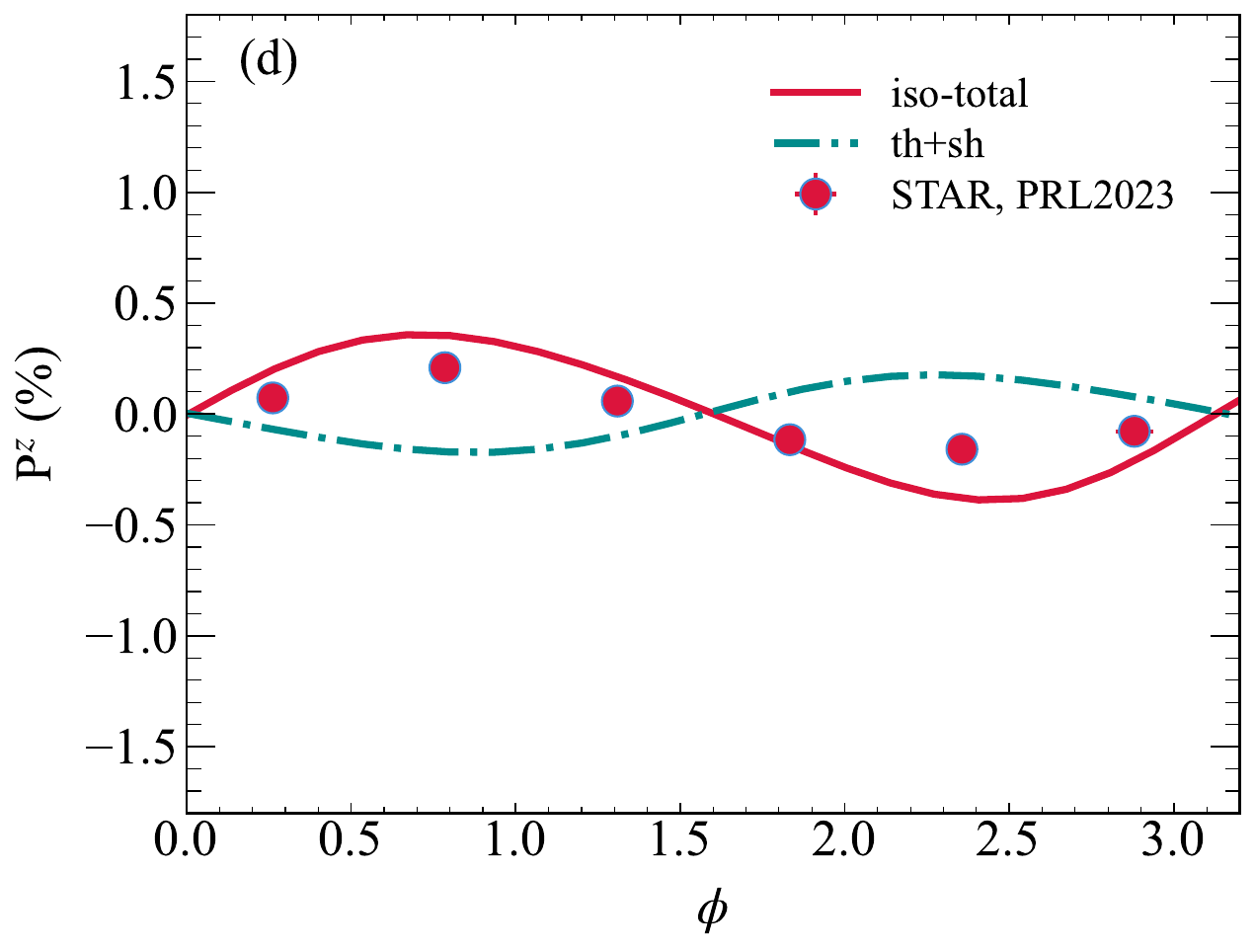}
\end{center}
\caption{(Color online) Comparison between the isothermal (`iso-total', solid) and
         standard thermal (`th+sh', dashed) polarization scenarios for $\Lambda$
         hyperons in 20--60\% Zr+Zr collisions at $\snn=200$~GeV
         ($p_T\in[0.5,3.0]$~GeV, $|\eta|<1$). (a) $-P^{y}$ vs.\ centrality. (b)
         $-P^{y}$ vs.\ $p_T$. (c) $-P^{y}$ vs.\ $\eta$. (d) $P_z$ vs.\ $\phi-\Psi_2$.
         STAR data are from Refs.~\cite{STAR:2025dgs,STAR:2023eck}.}
\label{f:iso_vs_thermal}
\end{figure*}

In summary, the \trento\ + CLVisc framework with the isothermal polarization scenario
captures the azimuthal dependence of $P_z$, including its negative sinusoidal phase and
overall amplitude, in reasonable agreement with STAR data. However, the $p_T$
dependence of the modulation amplitude $\langle P_z\sin[2(\phi-\Psi_2)]\rangle$ is
significantly overpredicted at $p_T\gtrsim 1.2$~GeV. Systematic scans of $f_v$, $k_T$,
nuclear structure, and bulk viscosity reveal that none of these effects, within their
physically reasonable ranges, can resolve the high-$p_T$ discrepancy. This points to
the need for additional physical mechanisms---such as the modified
acceleration-gradient-induced vorticity term ($\mathcal{S}_{\text{accT}}^{\mu}$),
electromagnetic fields, or late-stage hadronic rescattering---which may contribute
destructively to the high-$p_T$ longitudinal polarization. These effects will be
explored in future extensions of the present framework.

\subsection{Comparison between isothermal and standard thermal polarization scenarios}
\label{sec:3-5}

As discussed in Sec.~\ref{v1section2}, the polarization pseudo-vector can be evaluated
in two distinct scenarios: the isothermal scenario
[Eqs.~(\ref{eq:S_thermal_iso})--(\ref{eq:S_shear_iso})], where the freeze-out
hypersurface is approximated by a constant-temperature surface and the kinematic
vorticity and shear tensors $\omega_{\alpha\beta}$, $\Xi_{\alpha\beta}$ are used, and
the standard thermal scenario [Eqs.~(\ref{eq:th_shear})], which retains the full
temperature gradients through the thermal vorticity and shear tensors
$\varpi_{\alpha\beta}$, $\xi_{\alpha\beta}$. All results presented in
Secs.~\ref{sec:3-1}--\ref{sec:3-4} have been obtained in the isothermal scenario. In
this section, we directly compare the two approaches to assess their respective
strengths and limitations in describing the available STAR data for Zr+Zr collisions.

Figure~\ref{f:iso_vs_thermal} presents a systematic comparison between the isothermal
(solid) and standard thermal (dashed) polarization results across the four principal
observables. The comparison reveals a nuanced picture: neither scenario is universally
superior; rather, each has its own domain of validity, and their differences point to
the underlying physics that still needs to be understood.

For the out-of-plane global polarization $-P^{y}$, the two scenarios yield comparable
results in the centrality and pseudorapidity dependences, shown in panels (a) and (c),
respectively. The centrality trend and the overall magnitude are similar, and both agree with the STAR measurements within uncertainties.
The $\eta$ dependence is also well reproduced by both, with only minor differences in
the exact shape. These similarities reflect the fact that the temperature gradients on
the freeze-out hypersurface contribute only modestly to the angular-averaged global
polarization at this relatively high collision energy.

A notable difference, however, appears in the $p_T$ dependence of $-P^{y}$ shown in panel (b).
The standard thermal scenario (`th+sh', dashed) lies systematically higher than the
isothermal result at low $p_T$. At intermediate $p_T$, the thermal scenario
develops a more pronounced dip structure. The isothermal scenario, by contrast, shows
a milder dip and a more gradual recovery. When compared to the STAR data, the `th+sh'
result appears to track the measured trend slightly better at both low and intermediate
$p_T$, although the limited statistical precision of the data precludes a definitive
discrimination. The difference between the two scenarios in $-P^{y}(p_T)$ originates
from the additional temperature-gradient terms present in the standard thermal
vorticity and shear tensors [Eq.~(\ref{eq:varpi_xi})], which modify the $p_T$
dependence of both the thermal vorticity and shear contributions relative to their
isothermal counterparts.

The most striking contrast between the two scenarios emerges in the longitudinal
polarization $P_z(\phi)$ shown in panel (d). The isothermal scenario (solid) produces a
negative sinusoidal modulation with an amplitude of $\sim 0.4\%$--$0.5\%$, in good
agreement with the STAR data~\cite{STAR:2023eck} at all measured azimuthal angles. The
standard thermal scenario (dashed), however, yields a positive sinusoidal modulation of
comparable amplitude---i.e., it predicts the opposite sign of the azimuthal modulation.
This sign reversal is a direct consequence of the temperature-gradient contributions to
the shear tensor $\xi_{\alpha\beta}$: in the thermal scenario, the $\nabla T$ terms
alter the relative weight of the shear-induced longitudinal polarization to the point
of reversing the net $P_z$ phase. The fact that the isothermal scenario correctly
captures the experimentally observed sign, while the thermal scenario does not,
suggests that the isothermal treatment better captures the measured azimuthal phase at
$\snn=200$~GeV, where the hadronization temperature is expected to be nearly constant
across the freeze-out hypersurface~\cite{Becattini:2021suc,Becattini:2021iol}.

Taken together, these comparisons highlight the current status of the polarization
theory in isobaric collisions. The isothermal scenario successfully describes the
azimuthal structure of $P_z$, the centrality and $\eta$ dependences of $-P^{y}$, and
the overall magnitude of global polarization. The standard thermal scenario shows a
modest advantage in reproducing the $p_T$ dependence of $-P^{y}$, but fails
qualitatively for $P_z(\phi)$. This suggests that while the isothermal approximation
captures the essential physics at high collision energies, a more complete treatment
that properly accounts for temperature inhomogeneities in $P_z$---without spoiling its
correct azimuthal phase---may be needed for a unified description of all observables.
This remains an important direction for future theoretical development.

\section{Summary and outlook}
\label{section4}

We have presented a comprehensive theoretical study of global and
azimuthal-angle-dependent $\Lambda$ polarization in isobaric
$^{96}_{40}$Zr+$^{96}_{40}$Zr collisions at $\snn=200$~GeV using the \trento\ initial
condition model coupled to the (3+1)-D viscous hydrodynamic code CLVisc. A longitudinal
flow velocity gradient, controlled by $f_v$, is introduced into \trento\ for the first
time, providing an essential source of initial vorticity in this symmetric isobaric
system where the average participant thickness imbalance vanishes. The isothermal
polarization scenario, employing kinematic vorticity and shear tensors on a
constant-temperature freeze-out hypersurface, provides the theoretical foundation for
all calculations.

The framework provides a quantitative theoretical description of the global $\Lambda$
polarization $-P^{y}$ measured by STAR in Zr+Zr collisions at RHIC, reproducing its
centrality, $p_T$, and $\eta$ dependences. The $p_T$ dependence reflects the
competition between the thermal vorticity contribution (decreasing with $p_T$) and the
shear contribution (increasing with $p_T$), with the latter being essential at
intermediate and high transverse momenta.

Systematic scans of initial-state parameters yield several new insights:
(1) $f_v$ controls the overall magnitude of $-P^{y}$ and the relative
      thermal-vorticity weight in $-P^{y}(p_T)$. The Bjorken limit $f_v=0$
      significantly underpredicts the data, while $f_v=0.10$ is favored. The $\eta$
      profile remains essentially flat for all $f_v$ values, with the overall magnitude
      increasing steadily and being dominated by the thermal-vorticity contribution. 
(2) $k_T$ provides a complementary constraint: it enhances the polarization across
      the $p_T$ range by amplifying the shear contribution through a more strongly
      tilted fireball, and it drives the $\eta$ profile toward a convex shape peaked at
      midrapidity. The Bayesian-calibrated value $k_T=0.33$~GeV is preferred.
(3) The five nuclear structure configurations from the STAR isobar blind analysis
      yield nearly indistinguishable polarization, supporting
      the use of the combined Zr+Zr and Ru+Ru data set within the present model
      uncertainties and indicating that polarization is not sensitive to nuclear
      deformations at the $\beta_2,\beta_3\sim0.1$ level.

The decomposition into Fourier coefficients $P_{y,\mathrm{c0}}$ and $P_{y,\mathrm{c2}}$
cleanly separates the two polarization mechanisms: $P_{y,\mathrm{c0}}$ is
vorticity-dominated, $P_{y,\mathrm{c2}}$ shear-driven. Our model provides a
simultaneous description of both coefficients measured by STAR. We propose the $p_T$
dependence of $P_{y,\mathrm{c2}}$ as a clean experimental observable for isolating
shear-induced polarization in future measurements.

For the longitudinal polarization $P_z$, the isothermal scenario correctly reproduces
its azimuthal modulation. The $\phi$-integrated $P_z$ amplitude is nearly insensitive
to both $f_v$ and $k_T$, indicating that $P_z$ is driven by the flow velocity field
rather than the geometric tilt---consistent with earlier hydrodynamic
studies~\cite{Alzhrani:2022dpi}. However, the $p_T$-differential modulation $\langle
P_z\sin[2(\phi-\Psi_2)]\rangle$ shows some $f_v$ sensitivity at high $p_T$, providing a
complementary constraint on the longitudinal flow gradient. The persistent
overprediction of this observable at $p_T\gtrsim1.2$~GeV, unresolved by tuning $f_v$ or
$k_T$, points to additional mechanisms. Constant bulk viscosities ($\zeta/s$) provide
only a partial reduction, while the Duke parametrization enhances the modulation and
exacerbates the discrepancy, indicating that bulk viscosity alone cannot resolve the
high-$p_T$ tension.

The comparison between isothermal and standard thermal scenarios reveals that neither
achieves a unified description: the isothermal scenario captures the $P_z$ azimuthal
phase, while the thermal scenario offers modest improvement for $-P^{y}(p_T)$ but
predicts the wrong $P_z$ sign. These findings highlight the current theoretical
limitations and motivate the development of a more complete polarization framework, for
which the present work establishes a solid baseline.

\begin{acknowledgments}
We thank Cong Yi for useful comments. This work was supported by the National Natural
Science Foundation of China (Grant No.~12305138) and the Natural Science Foundation of
Hubei Province (Grant No.~2026AFB678). We acknowledge the use of DeepSeek-v4-pro for
language polishing, which helped us improve the clarity of the manuscript.
\end{acknowledgments}

\bibliographystyle{unsrt}
\bibliography{clv3}

\end{document}